\documentclass[11pt,reqno]{amsart}

\usepackage{cames0}
\usepackage[normalsize]{subfigure}
\usepackage[pdftex]{graphicx}
\usepackage[pdftex,colorlinks,bookmarksopen,bookmarksnumbered,citecolor=red,urlcolor=red]{hyperref}

\title[Homogenization of composites with interfacial debonding]%
      {Homogenization of composites with interfacial debonding using
        duality-based solver and micromechanics}

\author{P.~Gruber}
\address[P.~Gruber]{%
Department of Mechanics, Faculty of Civil Engineering, Czech Technical
University in Prague\\
Th\' akurova 7, 166 29 Prague 6, Czech Republic
}
\email{pavel.gruber@centrum.cz}
\urladdr{http://mech.fsv.cvut.cz/~grubepav}

\author{J.~Zeman}
\address[J.~Zeman]{%
Department of Mechanics, Faculty of Civil Engineering, Czech Technical
University in Prague\\
Th\' akurova 7, 166 29 Prague 6, Czech Republic
}
\email{zemanj@cml.fsv.cvut.cz}
\urladdr{http://mech.fsv.cvut.cz/~zemanj}

\author{J.~Kruis}
\address[J. Kruis]{%
Department of Mechanics, Faculty of Civil Engineering, Czech Technical
University in Prague\\
Th\' akurova 7, 166 29 Prague 6, Czech Republic
}
\email{jk@cml.fsv.cvut.cz}
\urladdr{http://mech.fsv.cvut.cz/web/people.php?id=23}

\author{M.~\v{S}ejnoha}
\address[M.~\v{S}ejnoha]{%
Department of Mechanics, Faculty of Civil Engineering, Czech Technical
University in Prague, Th\' akurova 7, 166 29 Prague 6, Czech Republic}
\address{
Centre for Integrated Design of Advanced Structures, Th\' akurova 7,
166 29 Prague 6, Czech Republic}
\email{sejnom@fsv.cvut.cz}
\urladdr{http://mech.fsv.cvut.cz/~sejnom}

\date{\today, Accepted in \sc Computer Assisted Mechanics and Engineering Sciences}

\newcommand{\reformulate}[1]{{\sf #1}}
\newcommand{\figname}[1]{\includegraphics{#1}}
\newcommand{\save}[1]{}

\newcommand{\setx}[1]{{\ensuremath{\Omega^{(#1)}}}}
\newcommand{\bset}{{\ensuremath{\Gamma}}}

\newcommand{\bsetj}[1]{{\ensuremath{\Gamma^{[#1]}}}}

\newcommand{\setuc}{{\ensuremath{\Omega^{\mathcal{UC}}}}}

\newcommand{\poissons}[1]{{\ensuremath{\nu^{(#1)}}}}

\newcommand{\abs}[1]{{\ensuremath{|#1|}}} % velikost oblasti
\renewcommand{\d}[1]{{\ensuremath{\mathrm{\,d}{#1}}}} % symbol diferenciálu
\newcommand{\average}[1]{{\ensuremath{\left\langle#1\right\rangle}}} % støední hodnota
\newcommand{\area}[1]{{\ensuremath{|#1|}}} % støední hodnota

\newcommand{\mj}[1]{{\ensuremath{\boldsymbol{\lambda}^{[#1]}}}}

\newcommand{\I}{{\ensuremath{\mathbf{I}}}}

\newcommand{\vek}[1]{\boldsymbol{#1}}
\newcommand{\mtrx}[1]{\mathbf{#1}}

\newcommand{\phs}[1]{^{(#1)}}

\newcommand{\stiffmateff}{\stiffmat^{\rm eff}}
\newcommand{\volfrac}{c}
\newcommand{\strainconcmat}{\mtrx{A}}
\newcommand{\pstressnconcmat}{\mtrx{T}}
\newcommand{\Pmat}{\mtrx{P}}

\newcommand{\MT}{{\rm MT}}
\newcommand{\tractionn}{\sigma}
\newcommand{\tractionnmax}{\sigma_{\max}}
\newcommand{\tractiont}{\tau}
\newcommand{\tractiontmax}{\tau_{\max}}

\newcommand{\Fref}[1]{\figurename~\ref{fig:#1}}
\newcommand{\Eref}[1]{Equation~\eqref{eq:#1}}
\newcommand{\Sref}[1]{Section~\ref{sec:#1}}
\newcommand{\Tref}[1]{\tablename~\ref{tab:#1}}

\newcommand{\PUC}{{\rm PUC}}
\newcommand{\FEM}{{\rm FEM}}
\newcommand{\BEM}{{\rm BEM}}
\newcommand{\FETI}{{\rm FETI}}

\newcommand{\myMatrix}[2]{%
\left[
\begin{array}{#1}%
#2%
\end{array}%
\right]
}

\newcommand{\objv}[1]{(#1)} % Volumetric object
\newcommand{\obji}[1]{[#1]} % Interfacial object
\newcommand{\clsr}[1]{\overline{#1}} % Closure

\newcommand{\puc}{\Omega}
\newcommand{\pucv}[1]{\puc^{\objv{#1}}}
\newcommand{\bpuc}{\Gamma}
\newcommand{\npuc}{\vek{n}}
\newcommand{\bpuci}[1]{\bpuc^{\obji{#1}}} % Internal boundaries
\newcommand{\nin}[1]{\vek{n}^{\obji{#1}}} % Internal normal
\newcommand{\y}{\vek{y}} % Position vector

\newcommand{\macron}{\vek{E}}
\newcommand{\macros}{\vek{\Sigma}}

\newcommand{\dmn}[1]{^{(#1)}}
\newcommand{\per}{^{*}}
\newcommand{\ipf}[1]{^{[#1]}}

\renewcommand{\u}{{\vek{u}}}
\newcommand{\micron}{\vek{\varepsilon}}
\newcommand{\micros}{\vek{\sigma}}

\newcommand{\trn}{{\sf ^T}}
\newcommand{\trct}{\vek{\lambda}}

\newcommand{\nmtrx}{\mtrx{\boldsymbol \nu}}
\newcommand{\pdmtrx}{\mtrx{\boldsymbol \partial}}

\newcommand{\stiffmat}{\mtrx{L}}
\newcommand{\test}[1]{\widehat{#1}}
\newcommand{\half}{{\textstyle \frac{1}{2}}}

\newcommand{\set}[1]{{\mathbb{#1}}} 
\newcommand{\dir}{_{\mathrm{D}}}
\newcommand{\n}{\vek{n}}
\renewcommand{\t}{\vek{t}}

\newcommand{\Nu}{\mtrx{N}_u}
\newcommand{\Bu}{\mtrx{B}_u}
\newcommand{\du}{\vek{d}_u}

\newcommand{\Nl}{\mtrx{N}_\lambda}
\newcommand{\dl}{\vek{d}_\lambda}

\newcommand{\K}{\mtrx{K}}
\newcommand{\f}{\vek{f}}
\newcommand{\Eb}{\mathcal{E}}

\newcommand{\pred}[1]{\tilde{#1}}

\newcommand{\R}{\mtrx{R}}
\newcommand{\dr}{\vek{d}_R}

\newcommand{\pinv}{{^\dag}}

\newcommand{\ind}{\hspace{2mm}}
\newcommand{\hind}{\hspace{4mm}}

\newcommand{\comm}[1]{{\bf #1}}
\newcommand{\an}[1]{\tiny \sf #1}
\newcommand{\loc}{_{\ell}}

\newcommand{\T}{\mtrx{T}}

\begin{document}\sloppy

\begin{abstract}
One of the key aspects governing the mechanical performance of
composite materials is debonding:~the local separation of reinforcing
constituents from matrix when the interfacial strength is exceeded. In
this contribution, two strategies to estimate the overall response of
particulate composites with rigid-brittle interfaces are
investigated. The first approach is based on a detailed numerical
representation of a composite microstructure. The resulting problem is
discretized using the Finite Element Tearing and Interconnecting
method, which, apart from computational efficiency, allows for an
accurate representation of interfacial tractions as well as mutual
inter-phase contact conditions. The candidate solver employs the
assumption of uniform fields within the composite estimated using the
Mori-Tanaka method. A set of representative numerical examples is
presented to assess the added value of the detailed numerical model
over the simplified micromechanics approach.
\end{abstract}

\maketitle

\section*{Keywords} 
%%%%%%%%%%%%%%%%%%%%
first-order homogenization; particulate composites; debonding;
\FETI~method; micromechanics

\section{Introduction}\label{sec:intro}
%%%%%%%%%%%%%%%%%%%%%%

Particle-reinforced composites, and the fibrous composites in
particular, present a progressive class of materials with a steadily
increasing importance in virtually all areas of structural
engineering, e.g.,~\cite{Cox:2006:QVT,Pendhari:2008:APC}. It is now
being generally accepted that one of the key factors governing the
mechanical performance of composite materials is the \emph{interfacial
  debonding}:~the partial separation of reinforcements from matrix
phase when the surface tractions locally exceed the interfacial
strength. Therefore, a considerable amount of research effort has been
invested into the development of a realistic model for
debonding-induced damage processes in heterogeneous media in the field
of materials science and engineering. Two major classes of models are
currently available: (i)~computational homogenization methods and
(ii)~micromechanics-based approaches. In spite of significant advances
in understanding the debonding phenomenon accomplished in recent
decades, both modelling approaches still suffer from certain
limitations.

The essential goal of the computational homogenization is to determine
the detailed distribution of local fields within a characteristic
heterogeneity pattern of the analyzed composites, usually represented
by a Periodic Unit Cell~(\PUC), employing a suitable discretization
technique. In the context of the debonding behavior, the most frequent
and versatile approach is based on the primal variant of the Finite
Element Method~(\FEM), e.g., \cite[and references
  therein]{Dobert:2000:NSID,Matous:2007:MMSP,Walter:1997:CMD} or a
mixed discretization approach such as the Voronoi Cell
\FEM~\cite{Li:2004:DCMM}. Alternative numerical schemes include the
Boundary Element Method~(\BEM)-based homogenization,
e.g.,~\cite{Chati:1998:PEP} or coupled
\BEM/\FEM~simulations~\cite{Gosz:1994:RITF}. The interfacial behavior
is, in all the previously mentioned cases, modeled in the form of a
traction-separation constitutive law, with the interfacial stiffness
as the basic material parameter. Such concept, however, inevitably
leads to problems for the perfect particle/matrix bonding, which
formally corresponds to the infinite value of interfacial
stiffness. In the actual implementation, of course, the perfect
bonding case is approximated using a large penalty-like term,
deteriorating the conditioning of the numerical problem manifested in
spurious traction oscillations and hence inaccurate prediction of
damage initiation, see
e.g.~\cite{Areias:2008:QSCP,Schellekens:1993:NIIE} and references
therein for additional discussion. The finite interfacial stiffness,
on the other hand, allows for mutual interpenetration of individual
constituents, resulting in a non-physical distribution of some of the
mechanical fields within composite. To circumvent these problems, an
Uzawa-type algorithm combined with the \FEM~or \BEM~discretization
were employed by Proch\'{a}zka and \v{S}ejnoha
in~\cite{Prochazka:2001:HLD,Prochazka:1995:DDR} to solve a single
fiber pull-out problem including interfacial Coloumb
friction. Moreover, the primal approaches of computational contact
mechanics were used with some success
in~\cite{Teng:2007:TSP,Wriggers:1998:CSID} to homogenize debonding
composites. All these studies, however, suffer from an increased
computational cost due to non-optimal convergence rates and as such
are not well-suited for the fully coupled FE$^2$ computational
homogenization~\cite{Feyel:2000:FE2,Kouznetsova:2001:AMMM}, where the
detailed \PUC~response serves as an ``effective'' non-linear
constitutive relation seen at the coarser scale of resolution.

The micromechanics-based solvers are, on the other hand, typically
based on a specific solution of field equations and limited
information of a microstructure, such as volume fractions of
individual phases. This makes the analytical approaches extremely
efficient from the computational point of view; their accuracy is,
however, to a major extent influenced by validity of the adopted
assumptions. Early contributions to the micromechanical modelling of
debonding composites include the contribution of
Benveniste~\cite{Benveniste:1985:EMB} and Pagano and
Tandon~\cite{Pagano:1990:MIB}, where the original uniform field
theories were extended to treat possible field discontinuities at the
fiber-matrix interphase. Additional improvements include introduction
of compliant interfaces with linear behavior~\cite{Hashin:1990:TPFC},
later extended for non-linear constitutive
relations~\cite{Dong:2000:MFE,Tan:2005:MTM}. In addition, layered
inclusion models combined with simple numerical procedures were
introduced in~\cite{Tandon:1996:ETM} to simulate specific
particle-reinforced composites. Several attempts have also been made
to increase the predictive capabilities of effective media models by
calibrating their response against the results of detailed numerical
simulations,
e.g.~\cite{Inglis:2007:CMDP,Rekik:2007:OEL,Sejnoha:1998:MEP}. However,
lacking reliable reference numerical solution, due to the reasons
summarized above, such a strategy can hardly reproduce the dominant
features of the complex interaction mechanisms in the material systems
under investigation.

In this contribution, we present an alternative numerical approach to
the numerical homogenization of debonding composites, which
circumvents the previously mentioned obstacles. The method itself
builds on the Finite Element Tearing and Interconnecting~(\FETI)
solver due to Farhat and Roux~\cite{Farhat:1991:FETI} and adopts
recent ideas in duality-based
solvers~\cite{Dostal:2007:FETI,Kruis:2008:RMI,Vlach:2001:MUD} to the
computational homogenization setting. The added value of the duality-
and \FETI -based framework can be summarized as follows:
\begin{itemize}

\item For the perfect complete bonding case, the method reduces to the
  original \FETI~algorithm; no artificial penalty-like stiffness is
  therefore needed to enforce the displacement continuity.

\item Since interfacial tractions are introduced independently from
  the displacement fields in the form of Lagrange multipliers, they
  are captured rather accurately in the whole loading
  regime. Therefore, the interfacial strength criteria can be directly
  adopted, cf.~\cite{Gruber:2008:FBH}.

\item In addition, the Lagrange multipliers allow us to efficiently
  treat frictionless contact conditions~\cite{Dostal:2009:QP}, making
  the numerical algorithm well-suited for providing reference
  solutions to micromechanics schemes.

\item As typical of \FETI -based methods, the dual problem involves
  the interface-related quantities only. This not only reduces the
  problem complexity by an order of magnitude, but also introduces a
  natural coarse level to the numerical
  algorithm~\cite{Dostal:2009:QP,Farhat:1994:OCP}, which opens the way
  to efficient and scalable iterative solvers.

\end{itemize}

In the rest of this work, these claims are clarified in more details
by addressing the fundamental case of composites with rigid-perfectly
brittle interfaces in the two-dimensional setting. The paper starts
with a brief overview of the first-order periodic homogenization of
composites with imperfect bonding of phases in \Sref{per_hmg}. The
numerical resolution of the unit cell problem using the \FETI~method
is covered in \Sref{feti_uc}, while a simple micromechanical model
based on the Mori-Tanaka method is introduced in
\Sref{micro_mt}. Performance of both approaches is compared in
\Sref{num_examples} for both regular as well as disordered real-world
microstructures. Finally, the most important results are summarized in
\Sref{concl}.

In the whole text, representation of the symmetric second and
fourth-order tensors of the Mandel type is systematically employed. In
particular, $a$, $\vek{a}$ and $\mtrx{a}$ denote a scalar value, a
vector or a matrix, respectively and the matrix representations are
defined, in the two-dimensional setting, via
\begin{eqnarray}
\vek{a} 
=
\myMatrix{c}{%
a_{11} \\ a_{22} \\ \sqrt{2} a_{12}}
,
&&
\mtrx{a}
=
\myMatrix{ccc}{%
a_{1111} & a_{1122} & \sqrt{2} a_{1112} \\
a_{1122} & a_{2222} & \sqrt{2} a_{2212} \\
\sqrt{2} a_{112} & \sqrt{2} a_{2212} & 2 a_{1212} \\
},
\end{eqnarray}
cf.~\cite[Section~2.3]{Milton:2002:TC}. Other symbols and
abbreviations are introduced in the text as needed.

\section{Overview of periodic homogenization}\label{sec:per_hmg}
%%%%%%%%%%%%%%%%%%%%%%%%%%%%%%%%%%%%%%%%%%%%%

In the current Section, the basic notation and essentials of the
first-order periodic homogenization theory are briefly summarized,
with the aim to provide a continuous setting compatible with the
\FETI-based discretization discussed next. For more detailed treatment
of individual steps, an interested reader is referred
to~\cite{Gruber:2008:HCM,Michel:1999:EPC}.

\subsection{Geometry of PUC and averaging}
%%%%%%%%%%%%%%%%%%%%%%%%%%%%%%%%%%%%%%%%%%
%
\begin{figure}[ht]
\hfill \subfigure[]{\figname{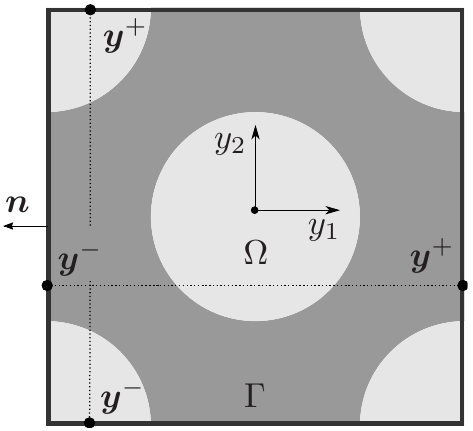}%
\label{fig:homo:decomp:uc}}
\hfill
\subfigure[]{\figname{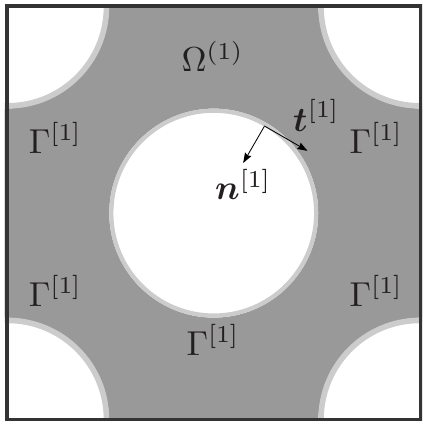}%
\label{fig:homo:decomp:matrix}}
\hfill
\subfigure[]{\figname{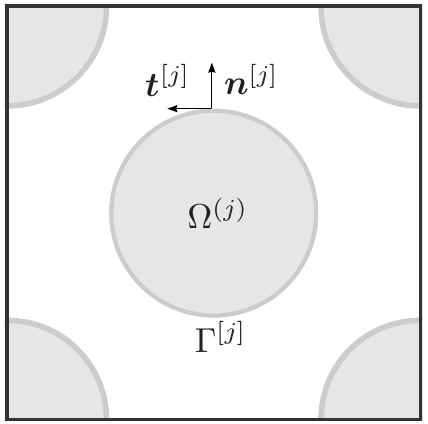}%
\label{fig:homo:decomp:inclusion}}
\caption{Decomposition of a \PUC; (a)~Unit cell, (b)~matrix
  phase, (c)~inclusion~(fiber).}
\label{fig:homo:decomp}
\end{figure}
Consider a Periodic Unit Cell~(\PUC) representing a cross-section of
composite material with long fibers,
see~\Fref{homo:decomp:uc}. Formally, the \PUC~is understood as an open
set $\puc \subset \set{R}^2$ with boundary $\bpuc$. Within the \PUC,
we distinguish $n$ disjoint sub-domains $\pucv{i}$. In the following
text, the index $i=1$ will be reserved for the multiply-connected
matrix phase, whereas simply-connected non-overlapping
heterogeneities, i.e. fibers, are enumerated by $i=2,3,\ldots, n$. It
will also prove useful to introduce the internal interfaces (denoted
by square brackets instead of round ones):
\begin{eqnarray}
\bpuci{j}
=
\clsr{\pucv{1}} \cap \clsr{\pucv{j}},
&&
\bsetj{1}
=
\bigcup_{j=2}^n\bsetj{j},
\label{eq:homo:set1}
\end{eqnarray}
i.e. the boundary $\bpuci{j}$ corresponds to the interface between
$j$-th fiber and the matrix phase, while $\bpuci{1}$ is reserved for
the whole internal matrix interface,
cf.~Figures~\ref{fig:homo:decomp:matrix}
and~\ref{fig:homo:decomp:inclusion}.\footnote{%
Note that in order to unify the notation, index $i$ consistently
ranges from $1$ to $n$, while $j \in \{ 2,3,\ldots, n \}$.}
The normal to the set $\puc$, defined on the whole boundary $\bpuc$,
is referred to as $\npuc$ with the appropriate adjustment for internal
boundaries. Note that for any $\y \in\bpuci{j}$, the associated
normals for the matrix phase and the $j$-fiber verify
\begin{equation}\label{eq:homo:oposite_normals}
\nin{1}(\y)=-\nin{j}(\y),
\end{equation}
cf.~Figures~\ref{fig:homo:decomp:matrix}
and~\ref{fig:homo:decomp:inclusion}.

Finally, we fix a nomenclature related to the determination of
averages for possibly discontinuous fields defined on \PUC. To that
end, consider a collection of functions $f\dmn{i}$ defined
independently on sub-domains $\pucv{i}$, jointly denoted as
\begin{equation}
f\left(\y\right)
=
\left\{
 \begin{array}{l@{\quad}l}
 f^{(1)}\left(\y\right),&\y\in\puc\dmn{1} \\
 f^{(2)}\left(\y\right),&\y\in\puc\dmn{2}\\
 \multicolumn{1}{c}{\vdots}\\
 f^{(n)}\left(\y\right),&\y\in\puc\dmn{n}
 \end{array}
\right.
.
\label{eq:homo:decomp_field_sets}
\end{equation}
Then, given two functions $f$ and $g$, we introduce the averaging
relation in the form, cf.~\cite[Section~2]{Michel:1999:EPC}:
\begin{equation}\label{eq:homo:def_average}
\average{\frac{\partial f(\y)}{\partial y_k} g(\y)} 
= 
\frac{1}{\area{\puc}}
\sum_{i=1}^n 
\left( 
\int_{\pucv{i}} \frac{\partial f\dmn{i}(\y)}{\partial y_k} g\dmn{i}(\y) \d\puc
-
\int_{\bpuci{i}} f^{[i]}(\y) g^{[i]}(\y) n^{[i]}_k( \y ) \d\bset
\right)
,
\end{equation}
where $\area{\puc}$ stands for the area of \PUC~and, in analogy
with~\Eref{homo:set1}, $f^{[i]}$ and $g^{[i]}$ denote the traces of
functions $f\dmn{i}$ and $g\dmn{i}$ on $\bpuci{i}$,
e.g.~\cite{rektorys:1999:VMI}. Note that for $\puc$-continuous
functions $f$~and~$g$, the second term of the right hand side
of~\eqref{eq:homo:def_average} vanishes due to
identity~\eqref{eq:homo:oposite_normals}, thus recovering the standard
format of the average on the unit cell.

\subsection{First-order homogenization framework}
%%%%%%%%%%%%%%%%%%%%%%%%%%%%%%%%%%%%%%%%%%%%%%%%%
%
In the current work, the \emph{strain controlled} approach to the
first-order computational homogenization~(in the terminology
introduced in~\cite{Michel:1999:EPC}) is adopted. Within this setting,
the analysis is split into two independent levels: the
\emph{microscale} corresponding to the behavior on the level of
individual constituents and the \emph{macroscale} providing the
overall response of the heterogeneous material under investigation. In
the sequel, we restrict our attention to the microscale problem and
consider a \PUC~subject to a \emph{macroscopic} strain $\macron$,
which, when combined with the governing equations of continuum
mechanics and the constitutive description of individual phases,
determines the distribution of \emph{microscopic} fields within the
\PUC. The average value of the microscopic stress $\micros$ then
yields the value of macroscopic stress $\macros$, implicitly defining
the \emph{homogenized} constitutive law.

Following this conceptual lead, we introduce an additive split of
displacement and strain fields $\u$ and $\micron$ in the form:
\begin{eqnarray}\label{eq:decomp}
\u\dmn{i}(\y) 
= 
\mtrx{X}(\y)\trn \macron
+
{\u\per}\dmn{i}(\y),
&&
\micron\dmn{i}(\y) 
=
\pdmtrx \u\dmn{i}(\y) 
=
\macron
+
{\micron\per}\dmn{i}(\y),
\end{eqnarray}
where the first part of the decompositions~\eqref{eq:decomp}
corresponds to an affine displacement field due to $\macron$, whereas
the second part is the correction due to the heterogeneity of the
\PUC. The matrices $\mtrx{X}$ and $\pdmtrx$ are defined in slightly
different form than in the standard approach~\cite{Bittnar:1996:NMM}
\begin{eqnarray}\label{eq:pos_der_def}
\mtrx{X}(\y)
= 
\myMatrix{cc}{
y_1 & 0 \\
0 & y_2 \\ 
\frac{1}{\sqrt{2}} y_2 & 
\frac{1}{\sqrt{2}} y_1
},
&&
\pdmtrx
= 
\myMatrix{cc}{
\frac{\partial}{\partial y_1} & 0 \\
0 & \frac{\partial}{\partial y_2} \\
\frac{1}{\sqrt{2}} \frac{\partial}{\partial y_2} &
\frac{1}{\sqrt{2}} \frac{\partial}{\partial y_1}
},
\end{eqnarray}
due to the adopted Mandel notation. 

In general, two conditions need to be satisfied to arrive at a
thermodynamically consistent macro-micro coupling. We start from the
macro-micro strain compatibility condition
\begin{equation}
\average{\micron(\y)} 
=
\macron, 
\end{equation}
transformed using~\Eref{decomp} and the integration
formula~\eqref{eq:homo:def_average} into a generalized boundary
conditions
\begin{eqnarray}\label{eq:strain_comp}
\average{\micron\per(\y)}
=
\vek{0}
& \Longleftrightarrow &
\oint_\bpuc 
\nmtrx(\y)
\u\per(\y) 
\d\bset
=
\vek{0},
\end{eqnarray}
imposed on the fluctuating displacement field $\u\per$. The $\nmtrx$
matrix appearing in~\eqref{eq:strain_comp} stores the components of
the normal vector and is defined analogously
to~\eqref{eq:pos_der_def}$_1$:
\begin{equation}\label{eq:nmtrx_def}
\nmtrx(\y) 
=
\myMatrix{cc}{%
n_1(\y) & 0 \\
0 & n_2(\y) \\
\frac{1}{\sqrt{2}} n_2(\y) & \frac{1}{\sqrt{2}} n_1(\y)
}.
\end{equation}

The second ingredient of the macro-micro scale transition is provided
by the Hill lemma
\begin{equation}\label{eq:homo:hill}
\macron\trn \macros
=
\average{\micron(\y)\trn \micros(\y)},
\end{equation}
imposing the equality of the work of macroscopic and microscopic
stresses and strains. Assuming a self-equilibrated microscopic stress
field (i.e. $\pdmtrx\trn\micros = \vek{0}$) and employing the
identity~\eqref{eq:homo:def_average}, the Hill lemma can be
transformed into
\begin{eqnarray}\label{eq:homo:hill_bound}
\average{{\micron\per}(\y)\trn \micros(\y) } = 0
& \Longleftrightarrow &
\oint_\bpuc 
{\u\per}(\y)\trn \trct(\y) 
\d\bset
= 0,
\end{eqnarray}
where $\trct$ denotes the traction vector defined via
\begin{equation}
\trct(\y)
=
\nmtrx(\y)\trn \micros(\y).
\end{equation}

Note that the two conditions~\eqref{eq:strain_comp}
and~\eqref{eq:homo:hill_bound} represent the \emph{necessary}
constraints for a consistent macro-micro scale tying. A particularly
convenient choice, especially for the adopted \emph{periodic} setting,
is provided by restricting the heterogeneous displacements $\u\per$ to
$\puc$-periodic fields. Then, for any homologous vectors $\y^+$ and
$\y^-$ located on the boundary $\bpuc$, cf.~\Fref{homo:decomp:uc}, the
periodicity and anti-periodicity of displacement and tractions holds:
\begin{eqnarray}\label{eq:homo:per_cond}
\u\per(\y^+) = \u\per({\y^-}),
&&
\trct(\y^+) = -\trct(\y^-),
\end{eqnarray}
which, together with the identity $\npuc(\y^+)=-\npuc(\y^-)$, ensures
the satisfaction of both kinematic and energetic consistency
conditions.

\subsection{Unit cell problem}
\begin{figure}[ht]
\hfill
 \subfigure[]{\figname{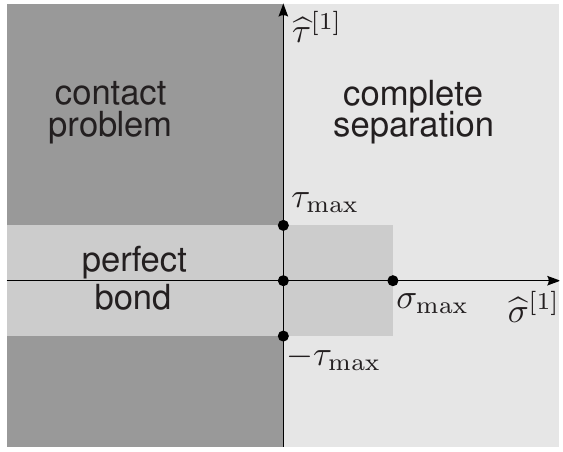}%
 \label{fig:interf_const_law}}
\hfill
 \subfigure[]{\figname{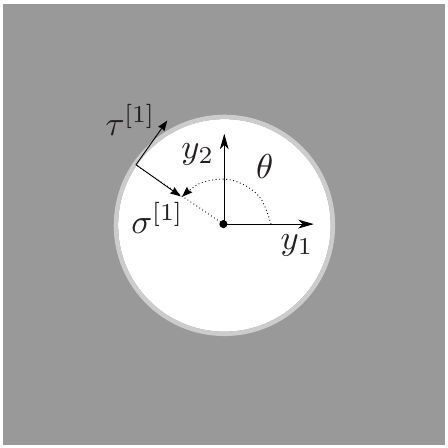}%
 \label{fig:inter_coord_syst}}
\hfill
\caption{Interfacial constitutive law; (a)~three-state model,
  (b)~local coordinate system.}
\label{fig:interf_const}
\end{figure}

%%%%%%%%%%%%%%%%%%%%%%%%%%%%%%
%
The distribution of the displacements and interfacial tractions
follows from the total energy functional in the form:
\begin{eqnarray}
\Pi\left(\test{\u}(\y), \test{\trct}(\y) \right)
=
\sum_{i=1}^{n}
\left( 
\half
\int_{\puc\dmn{i}} 
\test{\micron}\dmn{i}(\y)\trn
\stiffmat\dmn{i}
\test{\micron}\dmn{i}(\y)
\d\puc
-
\int_{\bpuc\ipf{i}}
\test{\trct}\loc\ipf{i}(\y)\trn
\test{\u}\loc\ipf{i}(\y)
\d\bpuc
\right)
\label{eq:feti:functional1},
\end{eqnarray}
where the symbol $\stiffmat\dmn{i}$ stands for the domain-wise
constant material stiffness matrix, $\test{a}$ is used to denote the
trial value of a quantity $a$; $\u\loc\ipf{i}$ and
$\trct\loc\ipf{i}(\y)$ correspond to displacement and traction fields
expressed in the local coordinate system as
\begin{eqnarray}
\myMatrix{c}{%
 u\ipf{i}_n(\y) \\
 u\ipf{i}_t(\y)
}
& = &
\myMatrix{c}{%
 u\ipf{i}_{1\ell}(\y) \\
 u\ipf{i}_{2\ell}(\y)
}
=
\T\ipf{i}(\y)
\myMatrix{c}{%
 u\ipf{i}_{1}(\y) \\
 u\ipf{i}_{2}(\y)
},
\\
\myMatrix{c}{%
 \tractionn\ipf{i}(\y) \\
 \tractiont\ipf{i}(\y)
}
& = &
\myMatrix{c}{%
 \lambda\ipf{i}_{1\ell}(\y) \\
 \lambda\ipf{i}_{2\ell}(\y)
}
= 
\T\ipf{i}(\y)
\myMatrix{c}{%
 \lambda\ipf{i}_{1}(\y) \\
 \lambda\ipf{i}_{2}(\y)
},
\label{eq:trct_loc}
\end{eqnarray}
with $\T\ipf{i}$ denoting the transformation matrix related to the
$i$-th interface:
\begin{equation}
\T\ipf{i}(\y)
=
\myMatrix{r}{%
 \nin{i}(\y)\trn \\
 \t\ipf{i}(\y)\trn
}
,
\end{equation}
cf.~Figures~\ref{fig:homo:decomp} and~\ref{fig:inter_coord_syst} for
an illustration. Employing the displacement and strain
decomposition~\eqref{eq:decomp} together with the interfacial
equilibrium conditions
\begin{equation}\label{eq:interf_equil}
\trct\ipf{1}(\y) = -\trct\ipf{j}(\y) 
\ \forall \y \in \bpuc\ipf{j}
\end{equation}
leads to a modified energy functional
\begin{eqnarray}
\Theta\left(
 \test{\u}\per(\y), \test{\trct}\ipf{1}(\y) 
\right)
& = &
\sum_{i=1}^{n}
\left(
\half
\int_{\puc\dmn{i}} 
{{\test{\micron}\per}{\dmn{i}}}(\y)\trn
\stiffmat\dmn{i}
{\test{\micron}\per}{\dmn{i}}(\y)
\d\puc
+
\int_{\puc\dmn{i}} 
{{\test{\micron}\per}{\dmn{i}}}(\y)\trn
\stiffmat\dmn{i}
\macron
\d\puc
\right)
\nonumber \\
& + &
\sum_{j=2}^{n}
\int_{\bpuc\ipf{j}}
\test{\trct}\loc\ipf{1}(\y)\trn
\left( 
 {\test{\u}\per\loc}{\ipf{j}}(\y) -
 {\test{\u}\per\loc}{\ipf{1}}(\y)
\right)
\d\bpuc
\label{eq:feti:functional2},
\end{eqnarray}
defined on the set of kinematically admissible displacements
\begin{equation}
\set{K}
=
\left\{
\test{\u}\per(\y) \mbox{ is } \puc\mbox{-periodic},
\test{\u}\per(\y) = \vek{0} \ \forall \y \in \bpuc\dir
\right\},
\end{equation}
with $\bpuc\dir$ denoting a part of the boundary, typically the corner
nodes of a \PUC, where the Dirichlet boundary conditions are imposed
to prevent the rigid body modes. The admissible interfacial tractions
are constrained to the statically admissible set $\Lambda$ discussed
in more detail in \Sref{iph_const_driver}. The ``true'' displacement
and traction fields then coincide with the saddle point of the
functional:
\begin{equation}\label{eq:uc_problem_def}
\left( \u\per(\y), \trct\ipf{1}(\y) \right)
=
\arg \max_{\test{\trct}\ipf{1}(\y) \in \Lambda} \min_{\test{\u}\per(\y) \in \set{K}} 
\Theta\left( \test{\u}\per(\y), \test{\trct}\ipf{1}(\y) \right),
\end{equation}
thereby defining the \emph{unit cell problem} for composites with
debonding phases.

\section{\FETI-based solution to unit cell problem}\label{sec:feti_uc}
%%%%%%%%%%%%%%%%%%%%%%%%%%%%%%%%%%%%%%%%%%%%%%%%%%
%
This Section presents an engineering approach to the solution of
non-linear debonding problem~\eqref{eq:uc_problem_def}. In particular,
a two-step iterative procedure is employed, where in the outer loop,
the static admissibility constraint $\test{\trct}\ipf{1} \in \Lambda$
is dropped to arrive at the prediction of traction distribution
$\pred{\trct}\ipf{1}$. Next, in the correction step, the trial
tractions are transformed into a physically admissible value using an
interfacial constitutive law and the loop is repeated until
convergence is achieved. It is fair to mention that, when compared to
more sophisticated techniques based on quadratic
programming~\cite{Dostal:2009:QP,Dostal:2007:FETI,Vlach:2001:MUD}, the
current technique is definitely less robust due to the missing
convergence theory. The definite advantage of the algorithm, on the
other hand, is its simplicity; as illustrated in the sequel, the
implementation systematically employs a standard \FETI~solver for
linear elasticity, see e.g.~\cite[Chapter~4]{Kruis:2006:DDM}.  In the
sequel, individual steps of the algorithm are summarized in some
detail.

\subsection{Prediction step}
%%%%%%%%%%%%%%%%%%%%%%%%%%%%
Following the basic idea of the \FETI-based
discretization~\cite{Farhat:1994:OCP,Kruis:2006:DDM}, the true fields
of fluctuating displacements and the corresponding strains are
approximated independently on individual domains $\puc\dmn{i}$
\begin{eqnarray}\label{eq:displ_discr}
\u\per{\dmn{i}}(\y) \approx \Nu\dmn{i}(\y) \du\dmn{i},
&
\micron\per{\dmn{i}}(\y) \approx \Bu\dmn{i}(\y) \du\dmn{i}
&
\forall \y \in \puc\dmn{i},
\end{eqnarray}
where $\Nu$ denotes the matrix of basis functions, $\Bu =
\mtrx{\partial} \Nu$ is the displacement-to-strain matrix and $\du$
stores the nodal displacements of individual components,
e.g.~\cite{Bittnar:1996:NMM}. The discretization of the interfacial
tractions in the predictor step is performed analogously:
\begin{eqnarray}\label{eq:trct_discr}
\pred{\trct}\ipf{1}(\y) \approx \Nl\ipf{1}(\y) \pred{\dl}\ipf{1}
&&
\forall \y \in \bpuc\ipf{1}.
\end{eqnarray}
Following the standard Ritz-Galerkin procedure, identical basis
functions are used for the discretization of the trial values.

After introducing the approximations~\eqref{eq:displ_discr}
and~\eqref{eq:trct_discr} into the energy
functional~\eqref{eq:feti:functional2}, the stationarity conditions
for $\du\dmn{i}$ and $\pred{\dl}\ipf{1}$ reduce to a system of linear
equations
\begin{eqnarray}
\K\dmn{i} \du\dmn{i} 
& = & 
\f\dmn{i} - \Eb\dmn{i}\trn \pred{\dl}\ipf{1},
\label{eq:feti:stac_cond_du} \\
\sum_{i=1}^n \Eb\dmn{i} \du\dmn{i}
& = & 
\vek{0}
\label{eq:feti:stac_cond_dm},
\end{eqnarray}
with the individual matrices expressed in the form:
\begin{eqnarray}
\K\dmn{i} 
& = & 
\int_{\puc\dmn{i}} 
 \Bu\dmn{i}(\y)\trn \stiffmat\dmn{i} \Bu\dmn{i}(\y) 
\d\puc,
\label{eq:feti:stiffmatrix}\\
\f\dmn{i}
& = & 
-\int_{\puc\dmn{i}} 
 \macron\trn \stiffmat\phs{i} \Bu\phs{i}(\y)
\d\puc,
\label{eq:feti:load}\\
\Eb\dmn{1}
& = &
-\sum_{j=2}^n \int_{\bpuc\ipf{j}}
 \T\ipf{1}(\y)\trn\Nl\ipf{1}(\y)\trn \T\ipf{1}(\y) \Nu\dmn{1}(\y)
\d\bpuc,
\label{eq:feti:L1} \\
\Eb\dmn{j}
& = &
\int_{\bpuc\ipf{j}}
 \T\ipf{1}(\y)\trn\Nl\ipf{1}(\y)\trn \T\ipf{1}(\y) \Nu\dmn{j}(\y)
\d\bpuc.
\label{eq:feti:Lj}
\end{eqnarray}
The solution of~\Eref{feti:stac_cond_du} exists only if and only if
\begin{equation}\label{eq:feti:solv_cond}
\R\dmn{i}\trn 
\left( 
 \f\dmn{i} - \Eb\dmn{i}\trn \pred{\dl}\ipf{1}
\right)
 = 
\vek{0},
\end{equation}
where $\mtrx{R}\dmn{i}$ stores the rigid body modes of the $i$-th
domain.

Note that in the actual implementation, the consistent compatibility
matrices $\Eb\dmn{i}$ appearing in Equations~\eqref{eq:feti:L1}
and~\eqref{eq:feti:Lj} are replaced with the lumped Boolean sparse
matrices, which enforce the displacement continuity discretely at the
corresponding nodes of the finite element mesh and furnish
$\pred{\dl}\ipf{1}$ with the physical meaning of the concentrated
nodal forces, see~\cite[Section~4.2]{Kruis:2006:DDM} for a more
detailed discussion.

Now we proceed with expressing the coefficients of fluctuating
displacements $\du\dmn{i}$ from the systems of equations
\eqref{eq:feti:stac_cond_du} in the form
\begin{equation}\label{eq:feti:expression_du}
\du\dmn{i}
=
\K\dmn{i}\pinv
\left( 
 \f\dmn{i} - \Eb\dmn{i}\trn \pred{\dl}\ipf{1}
\right)
+
\R\dmn{i} \dr\dmn{i}
\end{equation}
The first term in relation~\eqref{eq:feti:expression_du} corresponds
to the particular solution of the $i$-th component of the system
\eqref{eq:feti:stac_cond_du}, which is expressed using the generalized
inverse matrix $\K\dmn{i}\pinv$ replacing the inverse matrix for
singular $\K\dmn{i}$. The second term then corresponds to a
homogeneous solution, expressed as the linear combination of rigid
body modes $\R\dmn{i}$ with coefficients $\dr\dmn{i}$ to be
determined. Next, we substitute the coefficients of fluctuating
displacements $\du\dmn{i}$ from relations
\eqref{eq:feti:expression_du} to system~\eqref{eq:feti:stac_cond_dm}
and the solvability conditions~\eqref{eq:feti:solv_cond} to account again
for a potential singularity of matrices $\K\dmn{i}$:
\begin{eqnarray}
\sum_{i=1}^{n}\Eb\dmn{i} \K\dmn{i}\pinv \Eb\dmn{i}\trn \pred{\dl}\ipf{1}
-
\sum_{i=1}^{n}\Eb\dmn{i} \R\dmn{i} \dr\dmn{i}
& = &
\sum_{i=1}^{n}\Eb\dmn{i} \K\dmn{i}\pinv \f\dmn{i},
\label{eq:feti:dual_formulation} \\
\R\dmn{i}\trn 
\left( 
 \f\dmn{i} - \Eb\dmn{i}\trn \pred{\dl}\ipf{1}
\right)
& = & 
\vek{0}. 
\end{eqnarray}
Observe that the elimination of the primary unknowns $\du\dmn{i}$
leads to a substantially smaller dual problem, formulated in terms of
variables $\pred{\dl}\ipf{1}$ and $\dr\dmn{i}$. Such system can be
efficiently solved using the Modified Conjugate Gradient~(MCG) method,
augmented by the projection step to enforce the solvability
condition~\eqref{eq:feti:solv_cond}. As will be reported separately, a
special structure of the unit cell problem can be efficiently
exploited to construct the pseudoinverses $\K\dmn{i}\pinv$
efficiently. The robustness and scalability of the numerical algorithm
can further be extended by applying orthonormalization of matrices
$\R\dmn{i}$ and appropriate preconditioning procedure.

\subsection{Interfacial constitutive law}\label{sec:iph_const_driver}
%%%%%%%%%%%%%%%%%%%%%%%%%%%%%%%%%%%%%%%%%%%%
%
To close the problem statement, an appropriate interfacial
constitutive model needs to be specified. The particular choice
adopted in the current work neglects the interaction of the normal and
tangential strengths $\tractionnmax$ and $\tractiontmax$ and
accommodates the tangential slip in presence of compressive normal
stresses. Such assumptions lead to a three-state description appearing
in~\Fref{interf_const_law}. In particular, the following modes are
distinguished: (i)~perfect bond corresponding to an intact interface,
(ii)~frictionless contact state, where the local shear strength is
exceeded and sliding between fiber and matrix is activated and
(iii)~complete separation of phases.

The set of admissible tractions then specializes to
\begin{equation}\label{eq:stat_addm_def}
\Lambda 
= 
\left\{
\left( 
\test{\tractionn}\ipf{1}(\y) \leq \tractionnmax
\cap
| \test{\tractiont}\ipf{1}(\y) | \leq \tractiontmax
\right)
\cup
\left( 
\test{\tractionn}\ipf{1}(\y) \leq 0 \right)
\ \forall \y \in \bpuc\ipf{1}
\right\}.
\end{equation}
Correspondingly, the constraint $\test{\trct} \in \Lambda$ is simply
enforced by setting the traction value to zero whenever the
corresponding strength is exceeded by the predicted value
$\pred{\trct}$. Note that it follows from duality theory that the
condition $\test{\tractionn}\ipf{1} \leq 0$ is equivalent to the
non-interpenetration for the matrix and fibers~($u_n\ipf{j} \geq
u_n\ipf{1}$), cf.~\cite{Dostal:2007:FETI}.

\subsection{Implementation strategy}
%%%%%%%%%%%%%%%%%%%%%%%%%%%%%%%%%%%
%
The resulting incremental algorithm compatible with the adopted
interfacial law~\eqref{eq:stat_addm_def}, inspired by a related
study~\cite{Wriggers:1998:CSID}, is briefly summarized by means of a
pseudo-code shown in \Tref{code}. Note that in order to simplify the
exposition, a proportional loading path in the form
\begin{eqnarray}\label{eq:strain_path}
\macron( t ) = t \macron_{\max} 
& \mbox{with} &
0 \leq t \leq 1
\end{eqnarray}
is considered. Moreover, the irreversibility of debonding is enforced
using the static condensation~\cite[Section~2.5]{Bittnar:1996:NMM} of
the corresponding entries of $\dl\ipf{1}$, cf. lines 8 and 12
of~\Tref{code}.

\begin{table}[ht]
\small
\begin{tabular}{lll}
\hline
\an{i} & Initialize material parameters (phase elastic properties) and geometry (mesh) \\
       & \ind assemble stiffness matrices $\K\dmn{i}$, & Eqs.~(\ref{eq:feti:stiffmatrix})\\ 
       & \ind transformation matrices $\T\dmn{i}$ and compatibility matrices $\Eb\dmn{i}$ & Eqs.~(\ref{eq:feti:L1},\ref{eq:feti:Lj})\\
\an{ii} & Initialize loading data: macroscopic strain $\macron_{\max}$\\
       & \ind assemble load vectors $\vek{f}_{\max}\dmn{i}$ & Eqs.~(\ref{eq:feti:load})\\
\an{iii} & Computation FETI-based quantities: \\
       & \ind pseudoinverse $\K\dmn{i}\pinv$ and rigid body $\R\dmn{i}$ matrices\\
\an{iv}& Setup interfacial strengths and pseudo-times $t_1 < t_2 < \ldots < t_N$ \\       
\an{1} & \comm{for} $k=1:N$ (cycle through time steps)\\
\an{2} & \ind set $\f\dmn{i} = t_k \vek{f}_{\max}\dmn{i}$ and $\macron = t_k \macron_{\max}$ \\
\an{3} & \ind \comm{do} (outer loop) \\
\an{4} & \hind predict distribution of interfacial tractions $\pred{\dl}\ipf{1}$ using MCG
       & Eqs.~(\ref{eq:feti:dual_formulation},\ref{eq:feti:solv_cond})\\   
\an{5} & \hind \comm{for} $n=1:M$~(internal cycle though interfacial nodes)\\
\an{6} & \hind \ind extract $\pred{\tractionn}$ and $\pred{\tractiont}$ for $n$-th node\\
\an{7}& \hind \ind \comm{if} $\pred{\tractionn} \leq 0$ and $|\pred{\tractiont}| \geq \tractiontmax$~(contact problem) \\
\an{8}& \hind \hind set $\tractionn = \pred{\tractionn}$ and $\tractiont = 0$ by static condensation of $\Eb\dmn{i}$~(without possibility of "healing")\\
\an{9}& \hind \ind \comm{elseif} $\pred{\tractionn} \leq \tractionnmax$ and $|\pred{\tractiont}| \leq \tractiontmax$~(perfect bond)\\
\an{10}& \hind \hind set $\tractionn = \pred{\tractionn}$ and $\tractiont = \pred{\tractiont}$ \\
\an{11}& \hind \ind \comm{else} (complete separation) \\
\an{12}& \hind \hind set $\tractionn = 0$ and $\tractiont = 0$ by static condensation of $\Eb\dmn{i}$ (without possibility of "healing")\\
\an{13}& \hind \ind \comm{end}$_\mathrm{if}$ \\
\an{14}& \hind \comm{end}$_\mathrm{for}$\\
\an{15}& \ind \comm{until} interfacial conditions remain unchanged\\
\an{16}& \comm{end}$_\mathrm{for}$\\
\hline
\end{tabular}
\caption{Conceptual implementation of FETI-based solver for debonding problem}
\label{tab:code}
\end{table}

\section{Micromechanical model}\label{sec:micro_mt}
%%%%%%%%%%%%%%%%%%%%%%%%%%%%%%%
%
In the current Section, a simplified micromechanics-based solution to
the debonding problem is constructed, in order to, at the same time,
verify the developed FETI-based algorithm as well as to assess the
added value of the the detailed numerical modelling. To that end, a
composite is treated as a two-phase material, consisting of a matrix
phase~(domain $\setx{1}$), occupied by $(n-1)$ indistinguishable
particles~(related to domains $\setx{2}, \setx{3}, \ldots, \setx{n}$
in the FETI-based method). Moreover, the available geometrical
information is reduced to the phase volume fractions $\volfrac\phs{1}$
and $\volfrac\phs{2}$, defined as
\begin{eqnarray}
\volfrac\phs{1} = \frac{\abs{\setx{1}}}{\abs{\setuc}},  
&& 
\volfrac\phs{2} = 1 - \volfrac\phs{1},
\end{eqnarray}
where, in accord with the rest of the current section, a symbol
$\bullet\phs{1}$ is reserved for a quantity related to the matrix
phase while $\bullet\phs{2}$ collectively describes particle-related
unknowns.

Employing the simplified data, the overall stiffness of a two-phase
composite reads
\begin{equation}
\stiffmateff
= 
\stiffmat\phs{1}
+
\volfrac\phs{2} 
\left( \stiffmat\phs{2} - \stiffmat\phs{1} \right)
\strainconcmat\phs{2},
\end{equation}
where $\stiffmat\phs{1}$ and $\stiffmat\phs{2}$ are the matrix and
particle stiffnesses, respectively, and $\strainconcmat\phs{2}$ stands
for a strain concentration factor, relating a macroscopic strain
$\macron$ to the average strain in a particle. It follows from the
Benveniste's reformulation~\cite{Benvensite:1987:MTM} of the original
Mori-Tanaka method~\cite{Mori:1973:MTM} that the strain concentration
factor can be accurately approximated as
\begin{equation}
\strainconcmat\phs{2}
\approx
\strainconcmat_{\MT}\phs{2}
=
\pstressnconcmat\phs{2}
\left( 
 \volfrac\phs{1} \I
 + 
 \volfrac\phs{2} \pstressnconcmat\phs{2}
\right)^{-1},
\end{equation}
with $\pstressnconcmat\phs{2}$ denoting the partial strain
concentration factor~(with a physical meaning of the average strain in
a particle due to prescribed average strain in the matrix), expressed
in the form
\begin{equation}\label{eq:partial_stress_conc}
\pstressnconcmat\phs{2}
=
\left( 
 \I 
 + 
 \Pmat\phs{2} 
  \left( 
   \stiffmat\phs{2} - \stiffmat\phs{1}
  \right)
\right)^{-1}.
\end{equation}
The term $\Pmat\phs{2}$ appearing in \Eref{partial_stress_conc} is the
Walpole matrix~\cite{Walpole:1969:OOM}, related to a response of an
isolated inclusion embedded in an infinite matrix phase, which, under
the assumption of circular fibers, isotropic matrix and plane strain
state, can be explicitly expressed as, cf. \cite{Walpole:1969:OOM}:
\begin{equation}
\Pmat\phs{2} 
= 
\frac{1+\poissons{1}}{8 E\phs{1} \left( \poissons{1} - 1 \right)}
\left[ 
 \begin{array}{rrr}
 8 \poissons{1} - 5 & 1 & 0 \\
 1 & 8 \poissons{1} - 5 & 0 \\
 0 & 0 & 32\poissons{1} - 24
 \end{array}
\right],
\end{equation}
where $E\phs{1}$ a $\poissons{1}$ are the Young modulus and the
Poisson ratio of the matrix phase.

When assuming a phase-wise uniform distribution of mechanical fields
within heterogeneities, the stress value in a particle can be
estimated as
\begin{equation}
\micros\phs{2}( \y )
\approx
\micros_{\MT}\phs{2} 
=
\stiffmat\phs{2} \strainconcmat_{\MT}\phs{2} \macron,
\end{equation}
and, employing the equilibrium conditions~\eqref{eq:interf_equil},
converted into the approximate value of boundary tractions at the
particle/matrix interface
\begin{equation}\label{eq:traction}
\mj{1}( \theta )
\approx
\mj{1}_{\MT}( \theta ) 
= 
\nmtrx\ipf{1}(\theta)\trn
\micros\phs{2},
\end{equation}
where $\theta$ is an angle parameterizing a position on the interface
introduced in \Fref{inter_coord_syst} and the stress-traction
transformation matrix $\nmtrx\ipf{1}$ is expressed
using~\Eref{nmtrx_def} with $\n=\n\ipf{1} = \myMatrix{cc}{ -\cos
  \theta & -\sin \theta }\trn$. The algorithmic implementation of the
micromechanics-based solver closely follows the procedure introduced
by \Tref{code}. The debonding mechanism is reduced to a two-state
description, with one state describing an undamaged composite, whereas
the second state corresponds to completely debonded particles. In
particular, the perfect bonding at the matrix/particle interface is
maintained until the normal or tangential tractions do not exceed the
local strengths,
\begin{equation}
\tractionn\ipf{1}_{\MT}(\theta) \leq \tractionnmax 
\mbox{ and } 
| \tractiont\ipf{1}_{\MT}(\theta) | \leq \tractiontmax 
\mbox{ for all }
0 \leq \theta < 2\pi,
\end{equation}
where $\tractionn\ipf{1}_{\MT}$ and $\tractiont\ipf{1}_{\MT}$ denote
the normal and tangential tractions determined from
\eqref{eq:traction} via relation~\eqref{eq:trct_loc} with $\t\ipf{1} =
\myMatrix{cc}{\sin \theta & -\cos\theta}\trn$. If the previous
condition is violated for at least one value of $\theta$ for a given
level of macroscopic strain, we irreversibly set $\stiffmat\phs{2} =
\mtrx{0}$.

\section{Numerical examples}\label{sec:num_examples}
%%%%%%%%%%%%%%%%%%%%%%%%%%%%

Performance assessment of the numerical homogenization and the
micromechanics-based solvers will be performed on the basis of two
\PUC s representing cross-sections typical of unidirectional
composites: a parametric hexagonal unit cell model analyzed
in~\Sref{hex} and a twenty-particle unit cell treated in~\Sref{20puc},
which was obtained in~\cite{Zeman:2001:NEE} by a careful analysis of a
graphite/polymer composite sample. All \FETI-based examples presented
bellow are determined using an in-house code derived from a {\sc
  Matlab} implementation of finite element solver introduced
in~\cite{Alberty:2002:MIFEM} with the {\sc DistMesh} code, developed
by Persson and Strang~\cite{Persson:2004:SMG}, used to generate finite
element meshes. The material properties of individual phases and
interface are taken from a related study~\cite{Inglis:2007:CMDP} and
appear in \Tref{mat_data}. The plane strain assumption is adopted to
reduce the analysis to a two-dimensional cross-section and the
displacement fields are discretized using constant strain triangular
elements with the characteristic element/fiber diameter ratio
approximately equal to $0.05$. In all reported numerical simulations,
the following discretization of the time interval $[0,1]$ was
considered. First, assuming perfectly bonded phases, the value $t^*$
related to the debonding onset was determined. Next, the post-peak
branch $(t^*, 1]$ was uniformly sampled using $N$ equisized
  intervals. An interested reader is referred
  to~\cite{Gruber:2008:HCM} for additional implementation-related
  details.

\begin{table}[ht]
\centering
\begin{tabular}{lr}
\hline
\multicolumn{2}{c}{\bfseries \itshape Matrix} \\
\hline
Young's modulus, $E\phs{1}$ & $1$~GPa \\
Poisson's ratio, $\nu\phs{1}$ & $0.4$ \\
\hline
\multicolumn{2}{c}{\bfseries \itshape Inclusion} \\
\hline 
Young's modulus, $E\phs{2}$ & $150$~GPa \\
Poisson's ratio, $\nu\phs{2}$ & $0.3$ \\
\hline
\multicolumn{2}{c}{\bfseries \itshape Interface} \\
\hline 
Normal strength, $\tractionnmax$ & $0.02$~GPa \\ 
Tangential strength, $\tractiontmax$ & $0.02$~GPa \\ 
\hline
\end{tabular}
\caption{Material data of composite system}
\label{tab:mat_data}
\end{table}

\subsection{Hexagonal packing of particles}\label{sec:hex}
%%%%%%%%%%%%%%%%%%%%%%%%%%%%%%%%%%%%%%%%%%%
%
As the first example, we consider a hexagonal \PUC~subject to a
bi-axial tensile/compressive strain load characterized by
$
\macron_{\max} = \myMatrix{ccc}{0.02 & -0.02 & 0}\trn,
$
recall \Eref{strain_path}. Stress paths for the $\MT$~scheme
correspond to a fine discretization of the overall time~($N=100$),
whereas the value $N=10$ was adopted for the \FETI-based
algorithm. Note that the coarse time stepping of the latter approach
is mainly related to a high sensitivity of the post peak branch to
inevitable geometrical imperfections of the finite element mesh. As
illustrated in \Fref{hex_results}, taking $N=100$ leads to a
non-symmetric debonding mode, resulting in a substantially different
distribution of local fields and overall response than for the
symmetric deformation mechanism. When the time discretization is set
to $N=10$, on the other hand, the larger value of the overall strain
increment suppresses the mesh-induced imperfections and the solution
symmetry is maintained.

The resulting macroscopic stress-strain relations obtained by the \MT~and
\FETI-based approaches for the fiber volume fraction
$\volfrac\phs{2}=50\%$ appear in \Fref{stress_strain_puc}. For
comparison, the ideal debonding mode, corresponding to sufficiently
small time increment and perfectly symmetric mesh, is indicated by
gray color.

\begin{figure}[t]
\centering
\begin{minipage}{.48\textwidth}
\subfigure[]{\includegraphics[width=\textwidth]{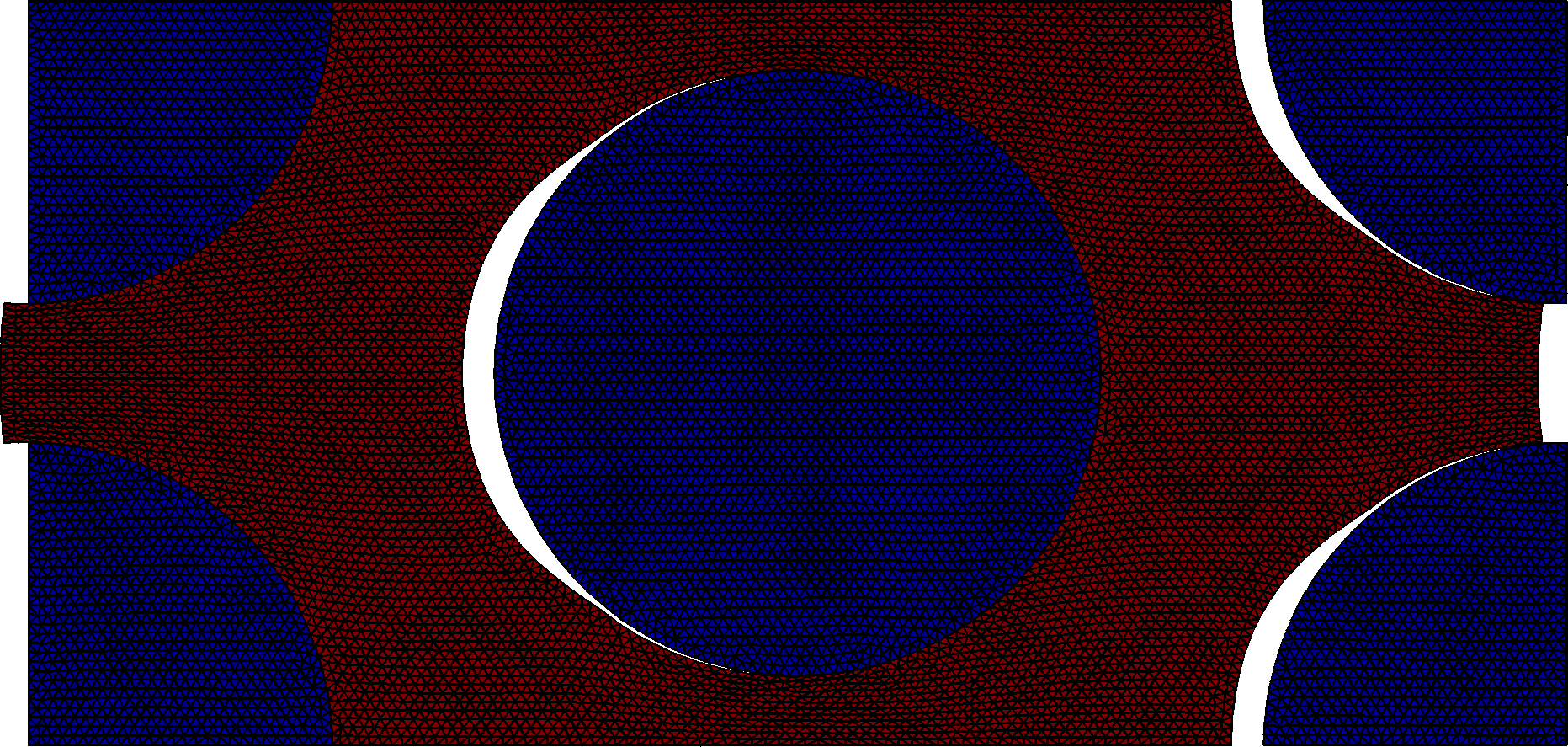}%
 \label{fig:hex_sym}}
\end{minipage}
\quad
\begin{minipage}{.48\textwidth}
\subfigure[]{\includegraphics[width=\textwidth]{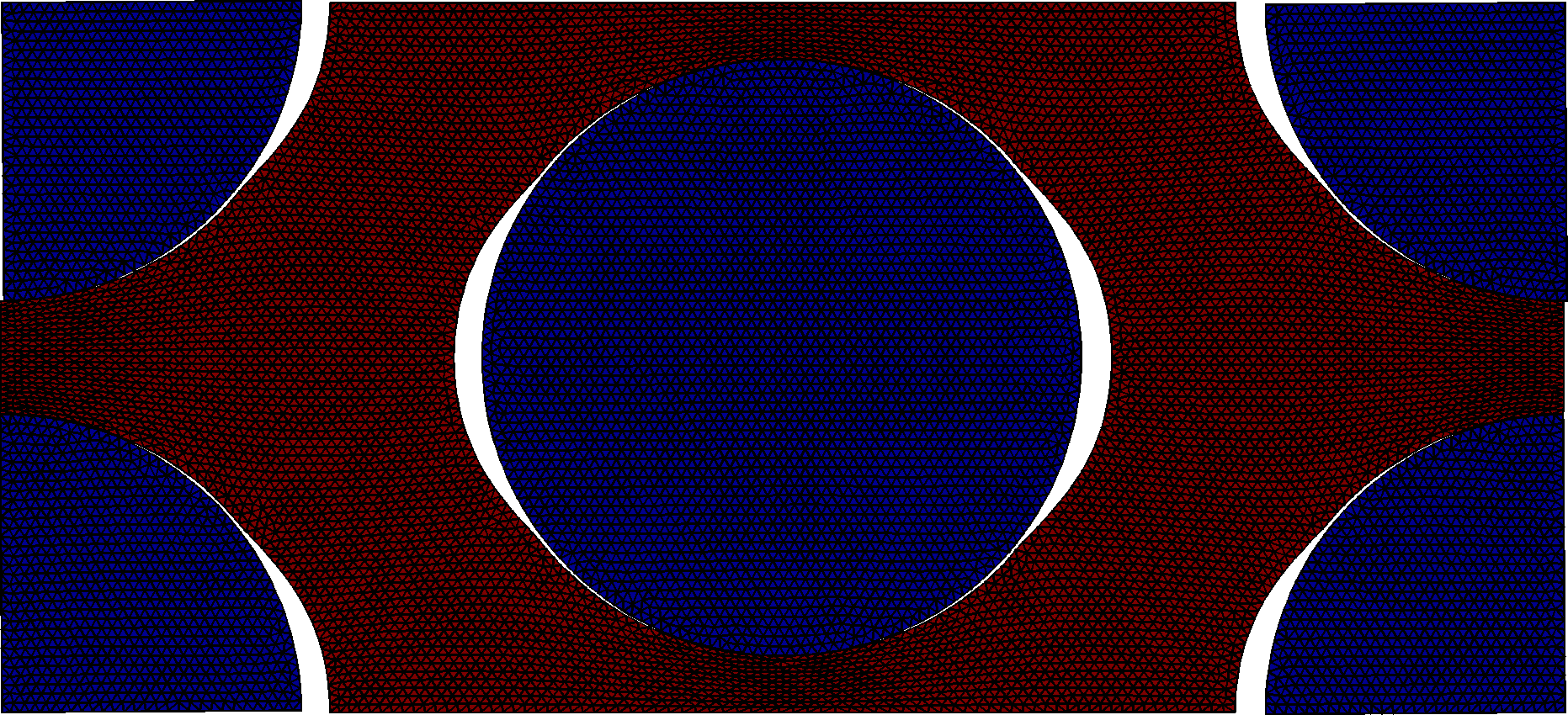}%
 \label{fig:hex_nonsym}}
\end{minipage}
\caption{Scheme of debonding initiation for a hexagonal \PUC; (a)~$N =
  100$, (b)~$N=10$. Displacements are scaled by factor of $20$.}
\label{fig:hex_results}
\end{figure}

\begin{figure}[ht]
 \includegraphics{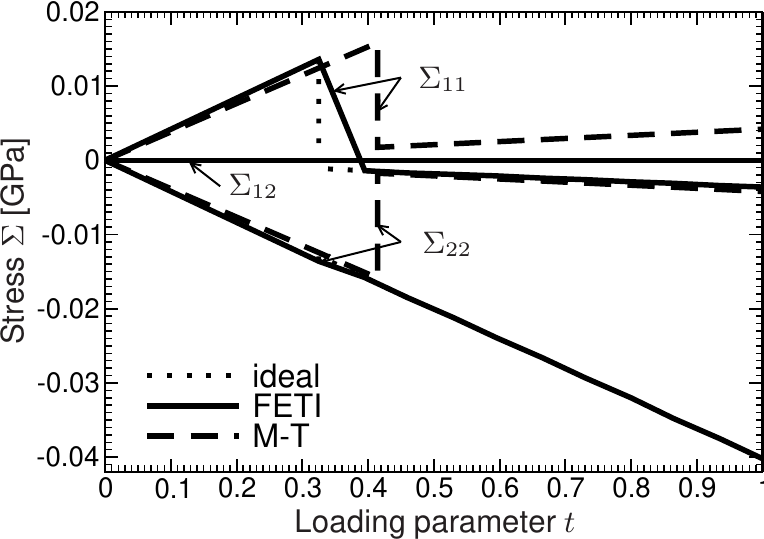}
 \caption{Macroscopic stress-strain diagram for hexagonal PUC;
   $\volfrac\phs{2} = 50\%$.}
 \label{fig:stress_strain_puc}
\end{figure}

In the elastic regime, both methods predict almost identical values of
macroscopic stresses with the MT~solver providing a slightly more
compliant response than the \FEM-based approach. Such behavior is
consistent with the fact that the Mori-Tanaka estimates, for material
systems with stiffer particles embedded in a weaker matrix phase,
coincide with the lower Hashin-Shtrikman variational bounds,
cf.~\cite{Castenada:1995:ESD}. A somewhat higher discrepancy is
observed for the prediction of a debonding onset $t^*$, where the
``unsafe'' MT result yields $t^*_{\MT} \doteq 0.41$, which by about
$25\%$ exceeds the reference value $t^*_{\FETI} \doteq 0.33$. The
difference can be attributed to the value of constant fiber stress
field adopted in the \MT~estimate. Such an assumption is
well-justified when determining the mean response of a material with
coherent interfaces, as demonstrated by excellent match of the elastic
branches in~\Fref{stress_strain_puc}, but becomes less accurate when
estimating the extreme stresses governing the failure process. This
claim is further supported by plots of interfacial tractions at the
debonding onset shown in \Fref{hex_traction_distrib}.\footnote{%
Note that the values of tractions shown in~\Fref{hex_traction_distrib}
correspond to different debonding times $t^*$ and therefore also to
different loading levels.}
We observe that, for the current case of $50\%$ particle volume
fraction, the deviation of the fiber stress from the assumed
uniformity in the \MT~method is further propagated to the values of
local tractions. It is also worth noting that, up to some
discretization-based effects, the \FETI~solution is free of spurious
traction oscillations appearing when introducing interface elements
between fibers and matrix as reported, e.g.,
in~\cite{Schellekens:1993:NIIE}.

\begin{figure}[ht]
\centering
\subfigure[]{\figname{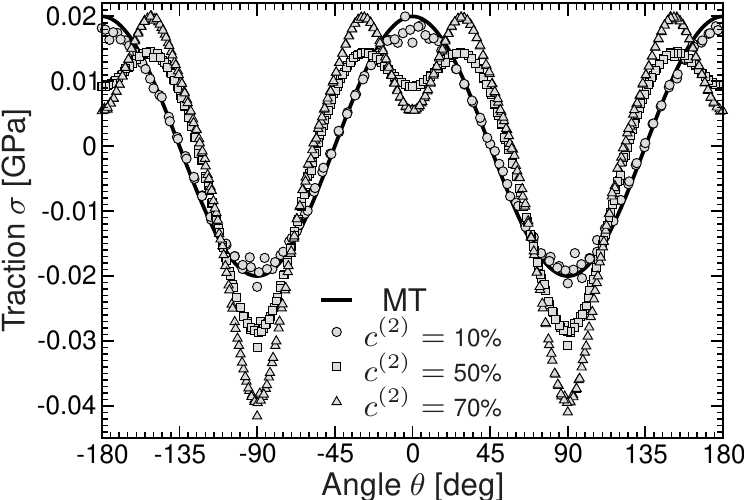}}
\quad
\subfigure[]{\figname{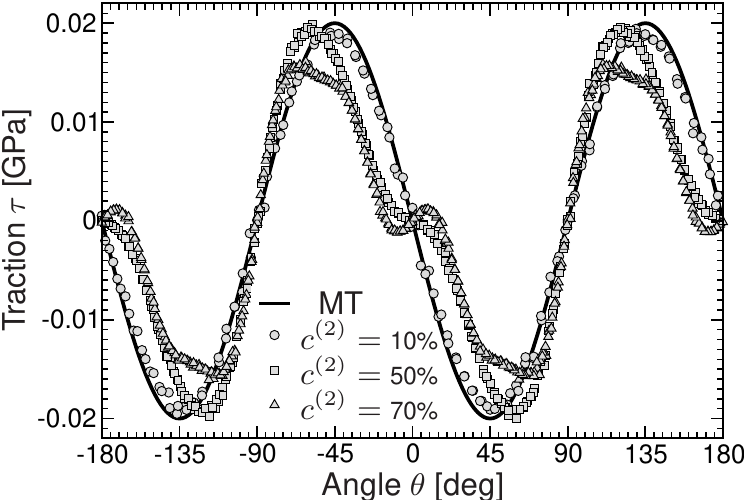}}
\caption{Distribution of interfactial tranctions at the onset of
  debonding; (a)~normal and (b)~tangential components; discrete values
  correspond to the~\FETI~solution, the~\MT~results are independent of
  $\volfrac\phs{2}$.}
\label{fig:hex_traction_distrib}
\end{figure}

After the debonding initiation, the discrepancies between both
approaches become even more pronounced. The \MT~solver predicts, due
to the assumption of complete interfacial failure, a sudden drop of
the stress values in both directions, reaching the values
corresponding to an isotropic porous medium in the post-peak
regime. The \FETI-based model, being capable of capturing partial
interfacial debonding, leads to a highly anisotropic mechanism: in the
compressive direction, the response remains almost insensitive to the
debonding phenomena, whereas the actual stress drop magnitude in the
$y_1$ direction exceeds the value predicted by the \MT~scheme, leading
eventually to the appearance of overall compressive stresses
$\Sigma_{11}$ due to intra-phase contact. As these conclusions remain
valid for both ``ideal'' and actual discretized stress paths, we will
perform the assessment on the basis of the latter results in the
sequel.

\begin{figure}[ht]
\centering
\subfigure[\FETI]{\figname{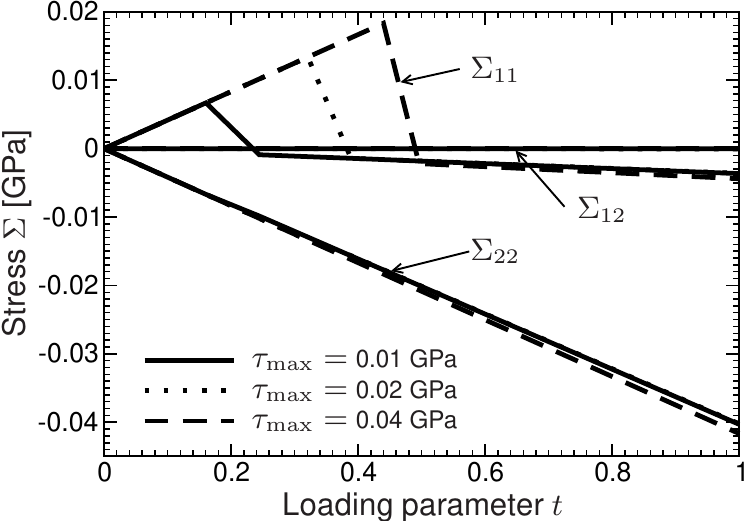}}
\quad
\subfigure[MT]{\figname{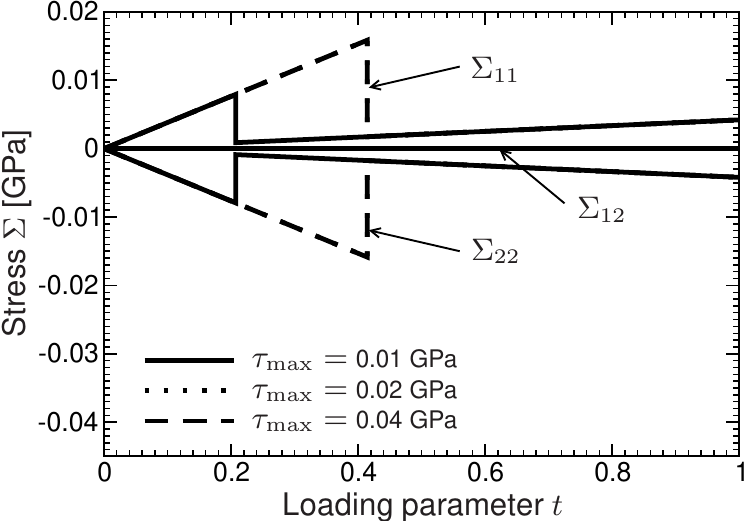}}
\caption{Influence of $\tractiontmax$ for hexagonal PUC;
  $\volfrac\phs{2} = 50\%$.}
\label{fig:hex_puc_tangential}
\end{figure}

Next, we investigate the influence of $\tractiontmax$ parameter on the
overall behavior of the hexagonal unit cell,
cf.~\Fref{hex_puc_tangential}. It appears that, for the current test,
the \MT~approach does not capture the change of failure mode when
increasing the interfacial strength from $\tractiontmax$ from $0.02$
to $0.04$~GPa. Such result can be again explained by inspecting
traction distribution for $\volfrac\phs{2} = 50\%$ shown
in~\Fref{hex_traction_distrib}. Due to the fact that $E_{11} = |
E_{22} |$, the \MT~theory predicts identical extereme values of normal
and shear tractions, which, in turn, implies that the debonding
initiation is governed solely by the minimum value of $\tractionnmax$
and $\tractiontmax$. The slight difference in the extreme shear and
tensile tractions predicted by the \FETI~solver, on the other hand, leads
to a three distict stress-strain curves, corresponding to a shear
initiation for $\tractiontmax = 0.01$~GPa and
$\tractiontmax=0.02$~GPa, while for $\tractiontmax=0.04$~GPa the
debonding starts in the normal mode. Note that similar results are
found when varying the $\tractionnmax$ parameter.

\save{
\begin{figure}[ht]
\centering
\subfigure[\FETI]{\figname{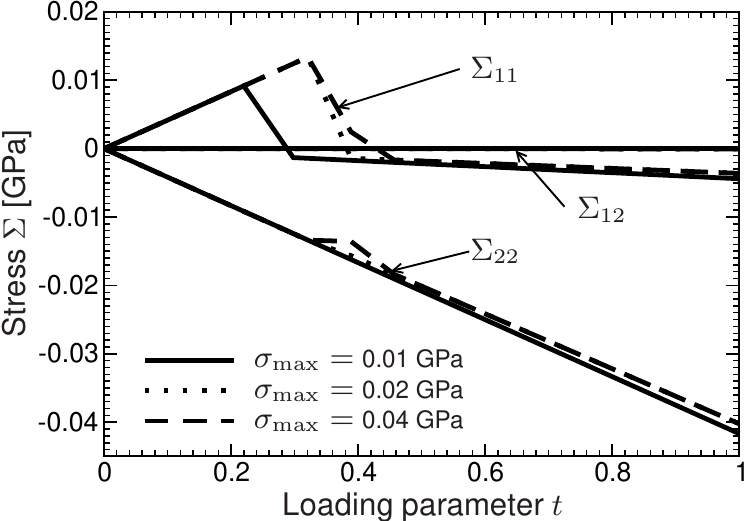}}
\quad
\subfigure[MT]{\figname{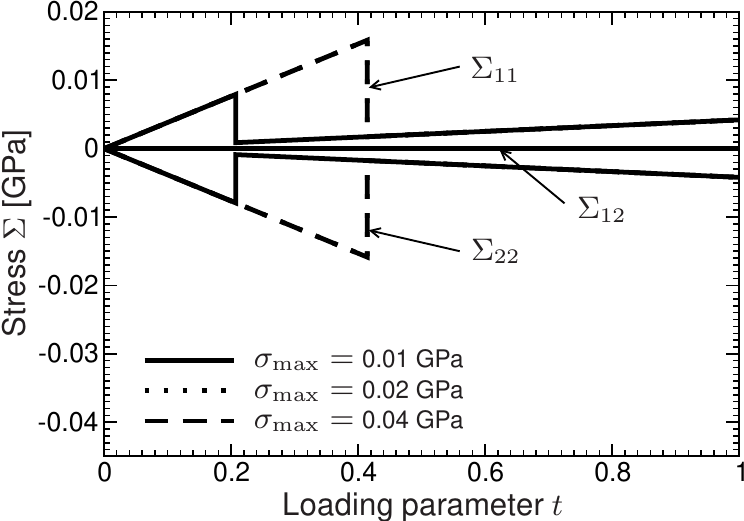}}
\caption{Influence of $\tractionnmax$ for hexagonal PUC;
  $\volfrac\phs{2} = 50\%$.}
\label{fig:hex_puc_normal}
\end{figure}

Results of a similar study obtained by varying the normal strength
$\tractionnmax$ are summarized in \Fref{hex_puc_normal}.
\reformulate{% 
Obviously, the \MT-based method is unable, at least for the current
load path, distinguish the influence from the case with variable
tangential strength.}
}

\begin{figure}[ht]
\centering
\subfigure[\FETI]{\figname{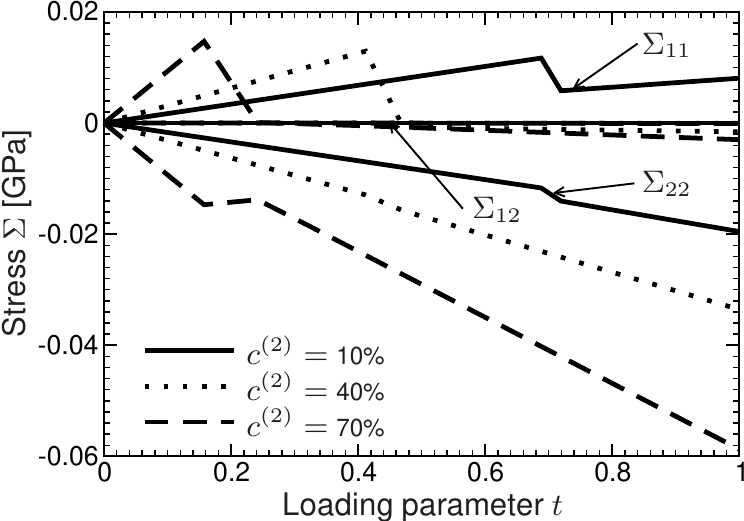}} 
\quad
\subfigure[MT]{\figname{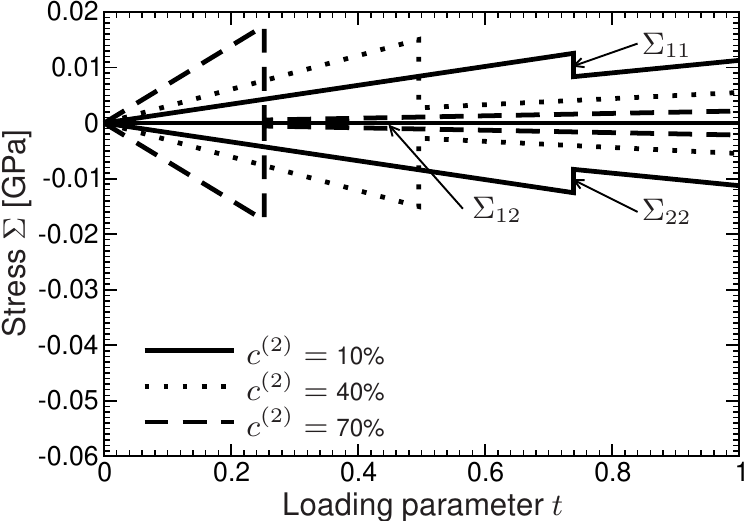}}
\caption{Influence of volume fractions for hexagonal PUC.}
\label{fig:hex_puc_volfrac}
\end{figure}

The last aspect to be analyzed remains the particle volume fractions,
cf.~\Fref{hex_puc_volfrac}. For the $\Sigma_{11}$ component, the
results of the \FETI-based solver predict a gradual transition of the
debonding branch from tensile to compressive stress values with
increasing particle volume fraction $\volfrac\phs{2}$. A closely
related trend, resulting from stress redistribution due to intra-phase
contact, is exhibited by the $\Sigma_{22}$ path, where the stress drop
after debonding changes its sign with increasing
$\volfrac\phs{2}$. The micromechanics-based solver, on the other hand,
leads to a symmetric response in terms of normal macroscopic stresses
with the difference in debonding time ranging from $7\%$ to
$40\%$. This trend is in a perfect agreement with traction plots
in~\Fref{hex_traction_distrib}, demonstrating that the mismatch
between traction distribution for the \MT~and \FETI~solvers generally
increases for densely packed particles due to more complex interaction
mechanism. In particular, for low volume fractions, the results of the
analytical approach almost exactly duplicate the numerical data; the
\MT~method, however, fails to reproduce the traction redistribution at
increased particle volume fractions $\volfrac\phs{2}$.

\subsection{Real-world microstructure}\label{sec:20puc}
%%%%%%%%%%%%%%%%%%%%%%%%%%%%%%%%%%%%%%
The universality of the conclusions reached for the hexagonal packing
of particles is further supported by an analogous analysis of a
twenty-fiber unit cell with particle volume fraction~$\volfrac\phs{2}
\doteq 43.9\%$, subject to a pure shear strain loading path with
$\macron_{\max} = \myMatrix{ccc}{0 & 0 & 0.02 \sqrt{2}}\trn$. The
mutual comparison of macroscopic stress path appears in~\Fref{20_puc},
storing the results of the \MT~method and, to illustrate the effect of
microstructural modeling, of the \FETI~solver for the disordered
microstructure and the regular hexagonal packing with identical fiber
volume fractions.

\begin{figure}[b]
\centering 
 \figname{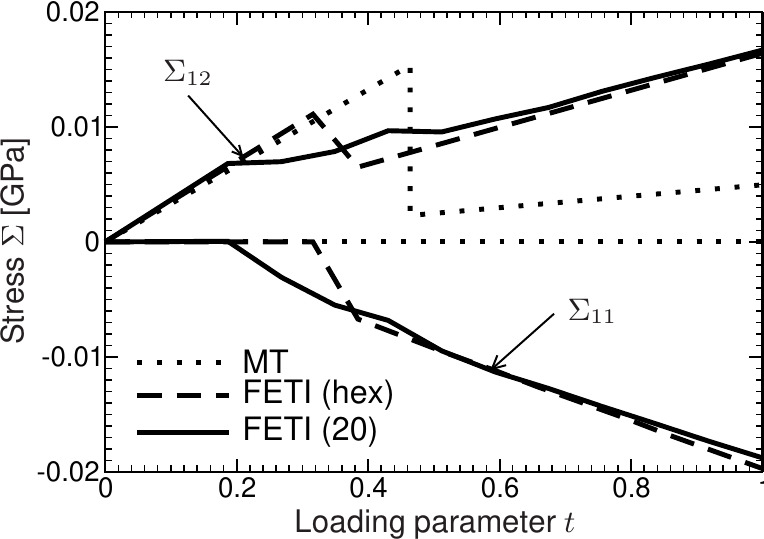}
\caption{Macroscopic stress-strain response for 20-fiber \PUC.}
\label{fig:20_puc}
\end{figure}

\begin{figure}[p]
\centering
\subfigure[$t \doteq 0.19$~(displacements are
    scaled by factor of~$20$)]{\includegraphics[height=48.6mm]{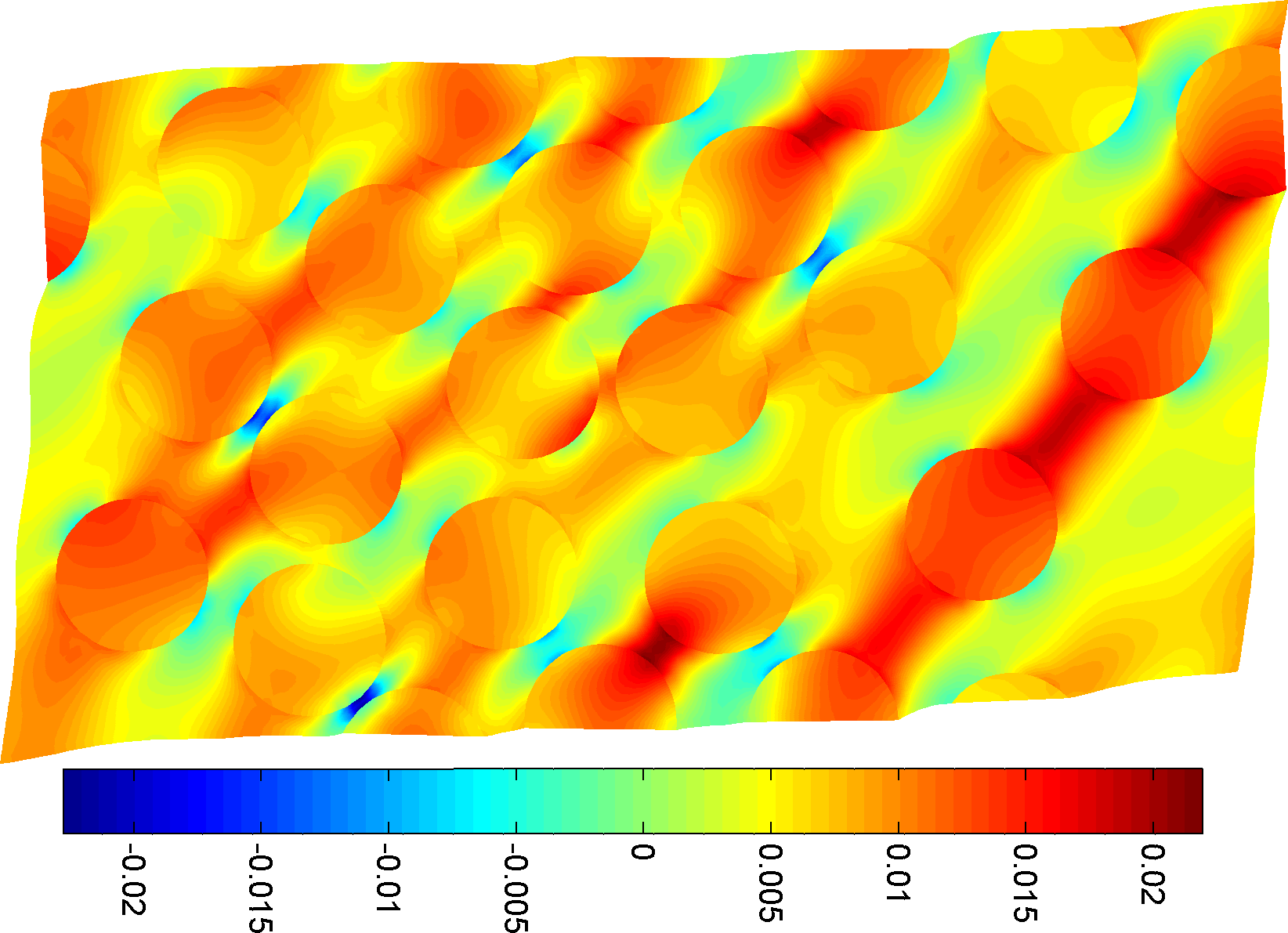}%
\label{fig:20_puc_stress_distrib_1}}
\quad
\subfigure[$t \doteq 0.32$~(displacements are
    scaled by factor of~$20$)]{\includegraphics[height=59.2mm]{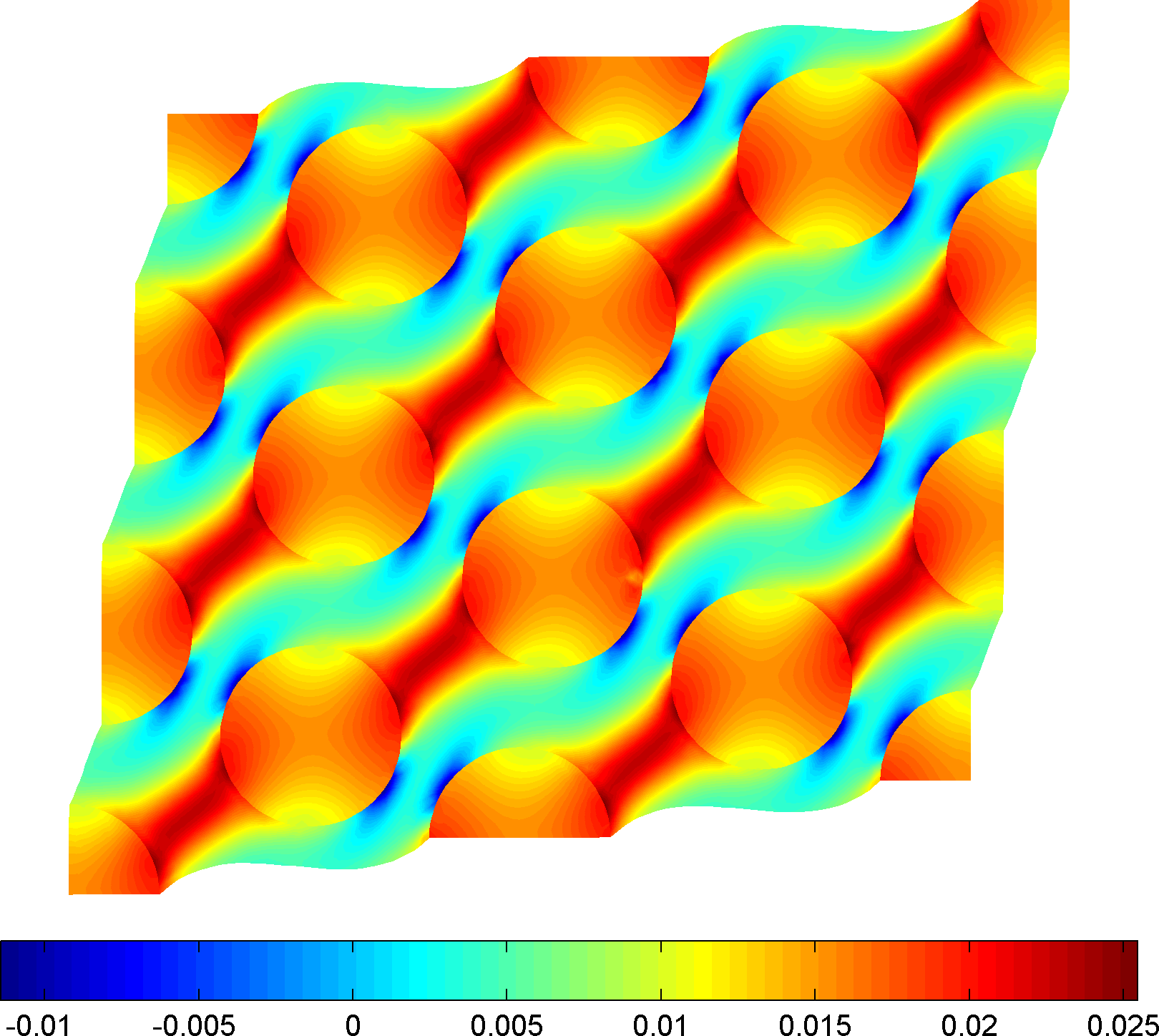}%
\label{fig:hex_puc_stress_distrib_1}}
\\
\subfigure[$t \doteq 0.27$~(displacements are
    scaled by factor of~$20$)]{\includegraphics[height=48.6mm]{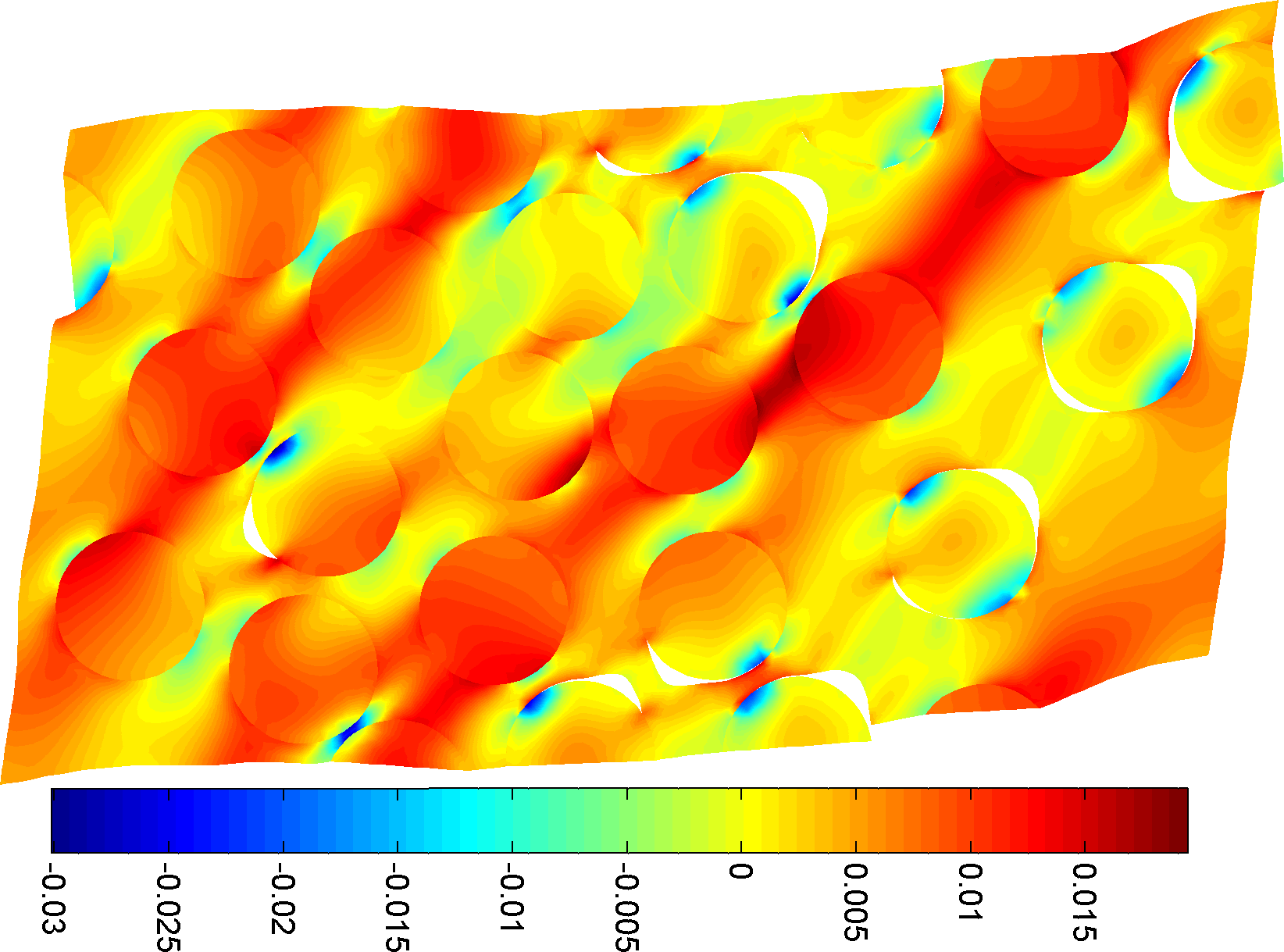}%
\label{fig:20_puc_stress_distrib_2}}
\quad
\subfigure[$t \doteq 0.39$~(displacements are
    scaled by factor of~$20$)]{\includegraphics[height=59.2mm]{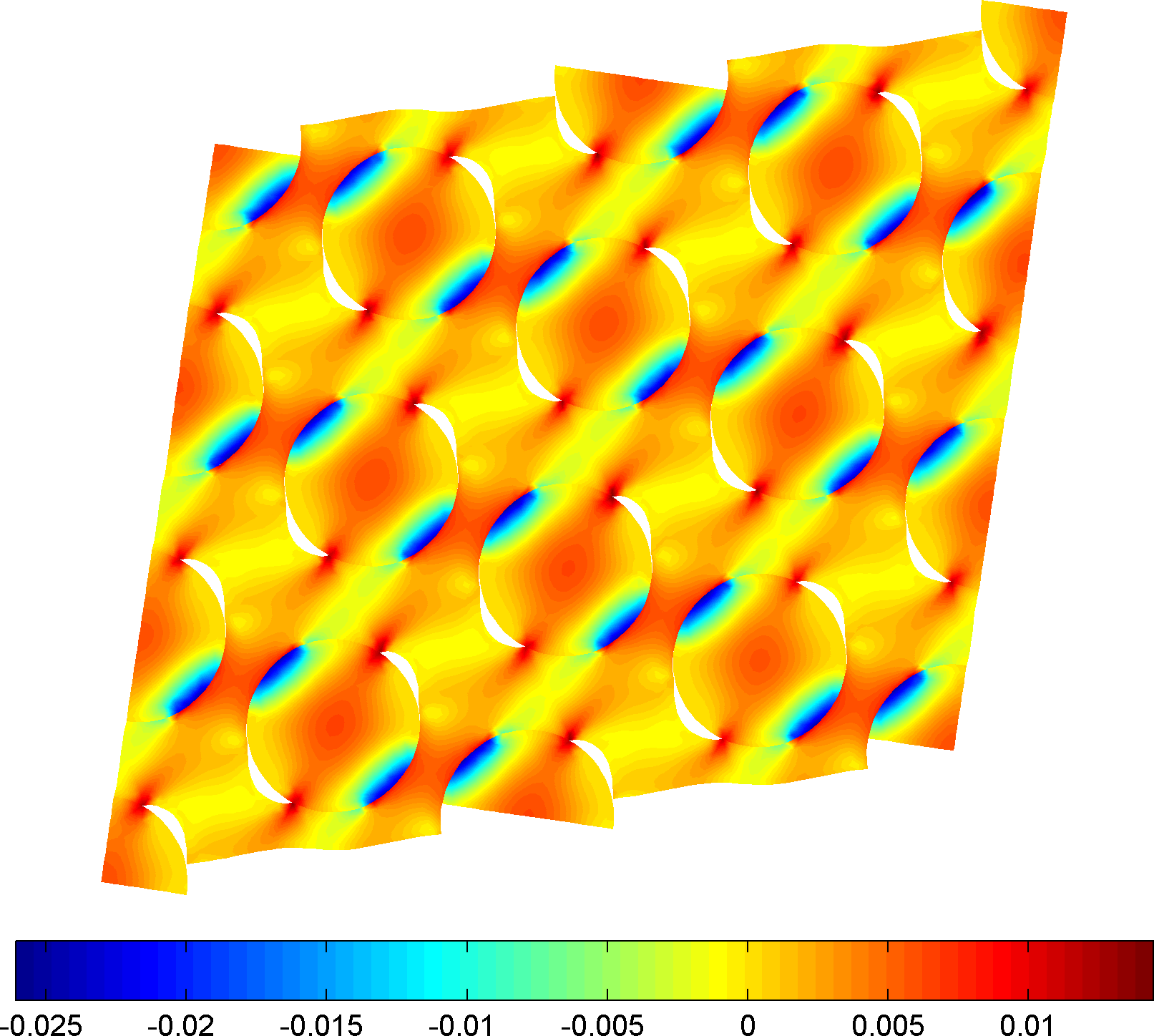}%
\label{fig:hex_puc_stress_distrib_2}}
\\
\subfigure[$t=1$~(displacements are
    scaled by factor of~$5$)]{\includegraphics[height=48.6mm]{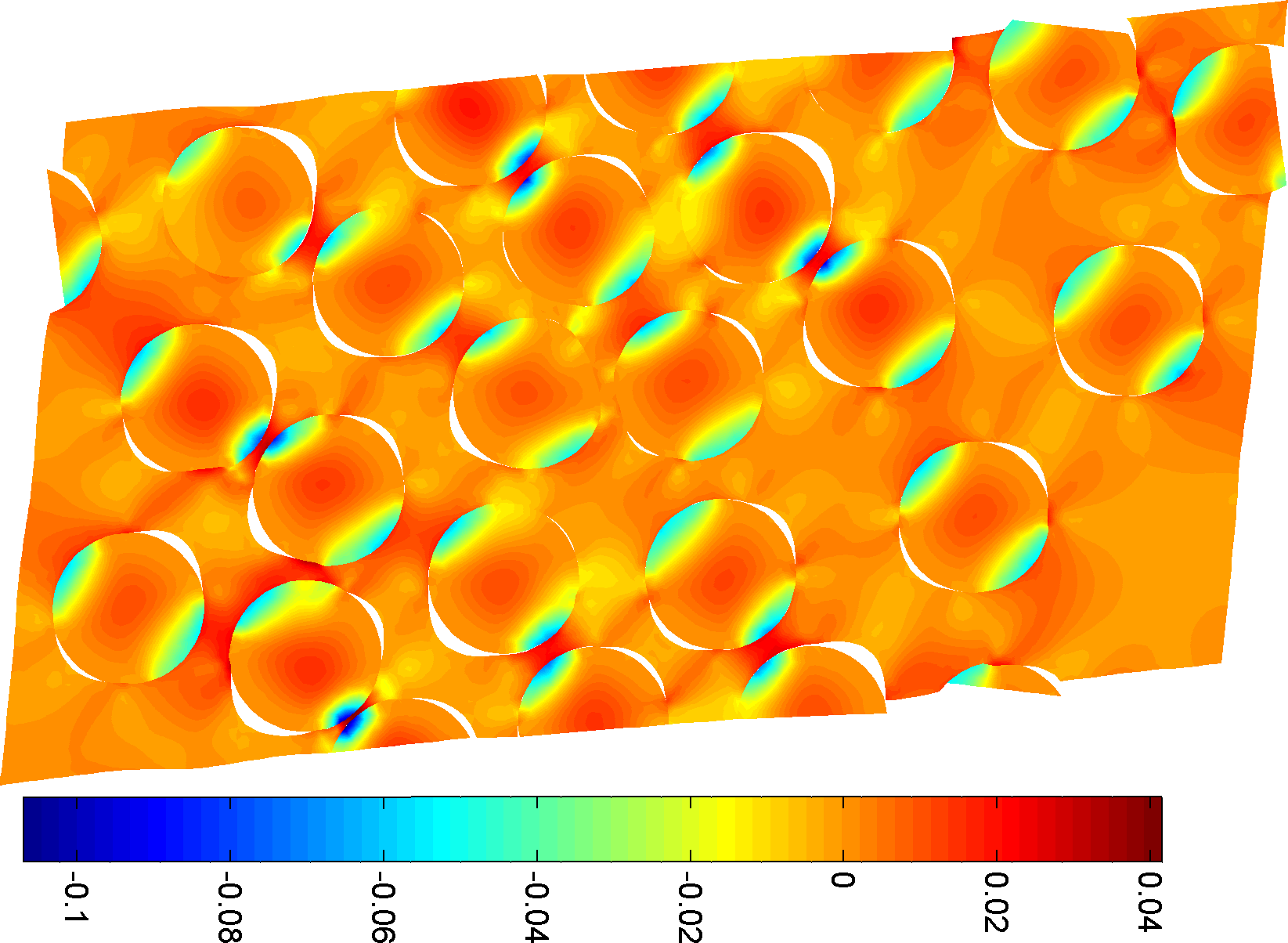}%
\label{fig:20_puc_stress_distrib_3}}
\quad
\subfigure[$t = 1$~(displacements are
    scaled by factor of~$5$)]{\includegraphics[height=59.2mm]{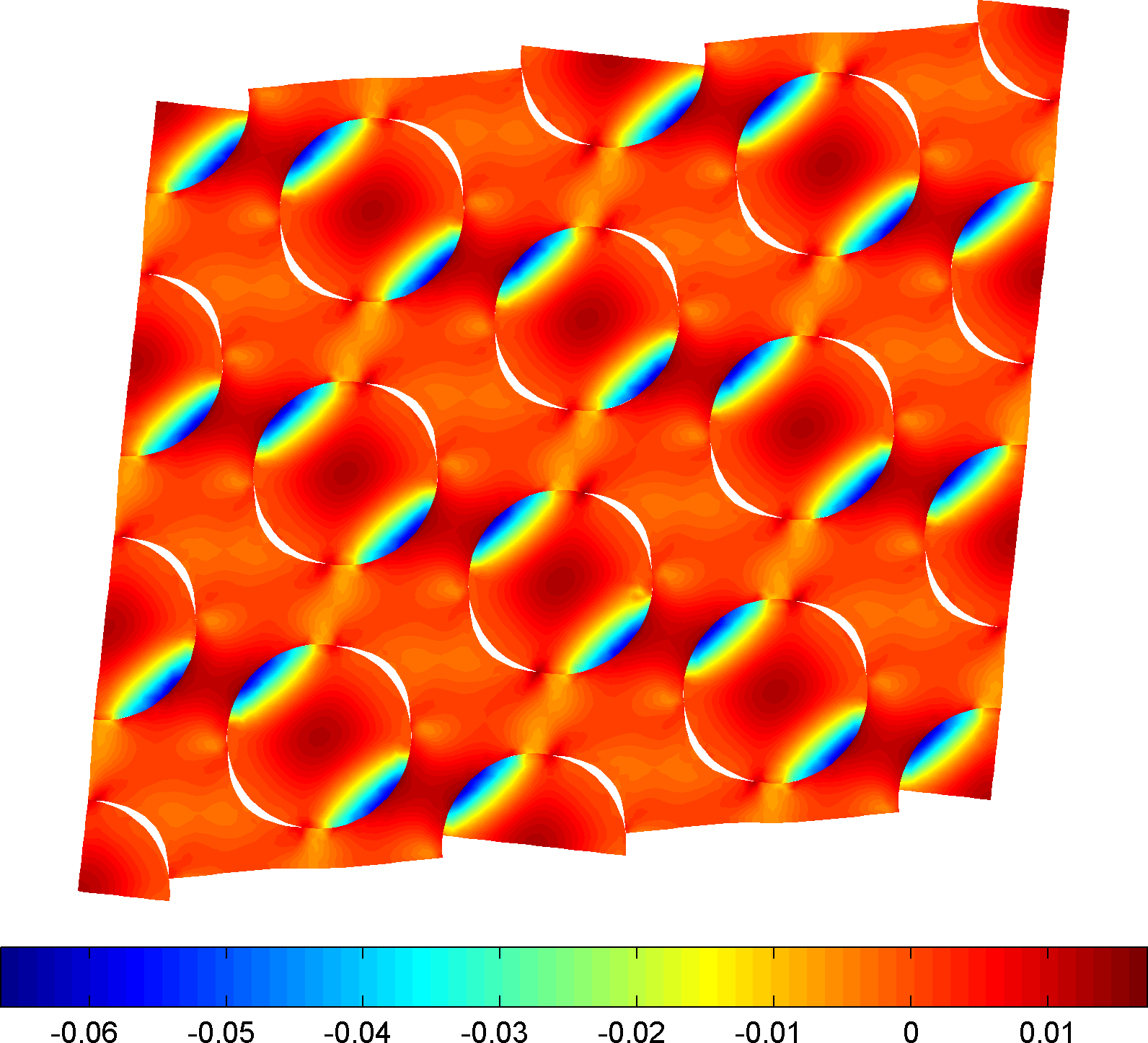}%
\label{fig:hex_puc_stress_distrib_3}}
\caption{Distribution of maximum principal stress for a
    20-fiber and equivalent hexagonal \PUC s. The values are given in GPa.}
\label{fig:20_puc_stress_distrib}
\end{figure}

The stress paths predicted by \FETI~reflect somewhat different failure
mechanisms for disordered and regular composite. In particular, the
inelastic path for the $20$-fiber unit cell exhibits a piecewise
linear evolution of the shear stresses, accompanied by a monotonically
descreasing normal stresses. The observed deformation mechanism can be
explained by the distribution of the local stress fields captured in
\Fref{20_puc_stress_distrib}. In particular, the debonding onset is
driven by stress amplification in the form of diagonal shear bands
visible in~\Fref{20_puc_stress_distrib_1}. After exceeding the
interfacial strengths, see~\Fref{20_puc_stress_distrib_2}, the
concentrations are released at newly formed discontinuity interfaces,
resulting in the formation of shifted shear bands connecting
non-debonded particles and in local matrix-fiber contacts. Such
mechanism is progressively repeated until the fully debonded state is
reached, cf.~\Fref{20_puc_stress_distrib_3}. For the regular
microstructure, the debonding initiates simultaneously at all fibers
and the stiffness immediately reaches the residual
value~(Figs.~\ref{fig:hex_puc_stress_distrib_1},
\ref{fig:hex_puc_stress_distrib_2} and
\ref{fig:hex_puc_stress_distrib_2}). It is worth noting that even the
debonded composites we observed~(up to minor discretization-induced
effects) the stress values $\Sigma_{11} = \Sigma_{22}$, which confirms
the spatial isotropy of the samples in question.

The proposed simple \MT-based solution, on the other hand, shows a
more compliant residual stiffness in the shear mode and predicts
appearance zero normal stresses. Such response is again a direct
consequence of the ``porous'' approximation adopted in the post-peak
regime. Finally observe that the debonding onset pseudo-times, as
predicted by different simulations, verify
$$
t^*_{\FETI(20)} \doteq 0.19 <
t^*_{\FETI(\mathrm{hex})} \doteq 0.32 <
t^*_{\MT} \doteq 0.46,
$$
cf.~\Fref{20_puc}, eventually demonstrating that the differences
between the regular and irregular microstructures due to local stress
amplicifactions can be well comparable with the value found when
comparing the analytical approach to the detailed numerical model.

\save{

seems to be
too simple to capture the appearance of non-zero normal contact
stresses. In addition, the failure onset prediction $t^*$ is
determined with rather large error~($\approx 150\%$), which is a
direct consequence of local stress amplifications due irregular
arrangement of particles visible from~\Fref{20_puc_stress_distrib_1}.
}
\save{%
Such results are further supported by \Fref{puc_stress_distrib},
showing the local distribution of the maximum principal stresses at
the debonding onset and for the fully developed debonding process,
respectively.
}

\section{Conclusions}\label{sec:concl}
%%%%%%%%%%%%%%%%%%%%%
In the current work, an overview of the \FETI-based procedure for the
homogenization of debonding composites was presented. In addition, the
developed algorithm was used to perform a detailed assessment of a
simplified micromechanics-based solver based on the \MT~method. The
results of the performed studies have led to the following
conclusions:

\begin{itemize}

\item The duality-based solver offers very promising approach to the
  homogenization of debonding composites, both from the point of view
  of numerical efficiency as well as the direct treatment of
  traction-based interfacial constitutive laws.

\item The simple micromechanics-based solver is fully capable of
  accurate prediction of the elastic part of the loading branch. The
  onset of debonding process is captured with the difference from the
  numerical results ranging from approximately $10\%$ to $40\%$ for
  regular packing of particles.

\item The assumption of completely debonded interfaces adopted in the
  \MT~solver, on the other hand, appear to be quite unrealistic and is
  mainly responsible for the inconsistent predictions in the post-peak
  regime.

\item The results presented in \Sref{20puc} clearly confirm that the
  apparent post-critical response is influenced not only by details of
  particle arrangement, but also by the \PUC~size,
  cf.~\cite{Sejnoha:2004:NVA}. When dealing with inherently random
  real world composites, the choice of an appropriate statistically
  equivalent \PUC~becomes rather delicate and generally requires a
  proper statistical characterization in terms of geometry and
  mechanical quantities of interest,
  see~\cite{Kanit:2003:DSRVE,Zeman:2007:FRM} for additional
  discussion.

\end{itemize}

The future extension of the \FETI-based solver would involve a
treatment of more realistic interfacial constitutive laws. For the
micromechanics-based solver, the essential feature to incorporate
seems to be the debonding-induced anisotropy in composite including
the effect of contact stresses. Both topics enjoy our current interest
and will be reported on separately.

\subsection*{Acknowledgements}
%%%%%%%%%%%%%%%%%%%%%%%%%%%%%%
Financial support of this work provided by the Grant Agency of the
Czech Republic, project no. GA\v{C}R~106/08/1379, is gratefully
acknowledged.


\begin{thebibliography}{10}

\bibitem{Alberty:2002:MIFEM}
J.~Alberty, C.~Carstensen, S.~A. Funken, and R.~Klose, \emph{{MATLAB}
  implementation of the finite element method in elasticity}, Computing
  \textbf{69} (2002), no.~3, 239--263.

\bibitem{Areias:2008:QSCP}
P.~M.~A. Areias and T.~Rabczuk, \emph{Quasi-static crack propagation in plane
  and plate structures using set-valued traction-separation laws},
  International Journal for Numerical Methods in Engineering \textbf{74}
  (2008), no.~3, 475--505.

\bibitem{Benveniste:1985:EMB}
Y.~Benveniste, \emph{The effective mechanical behaviour of composite materials
  with imperfect contact between the constituents}, Mechanics of Materials
  \textbf{4} (1985), no.~2, 197--208.

\bibitem{Benvensite:1987:MTM}
Y.~Benveniste, \emph{A new approach to the application of {M}ori-{T}anaka theory in
  composite materials}, Mechanics of Materials \textbf{6} (1987), no.~2,
  147--157.

\bibitem{Bittnar:1996:NMM}
Z.~Bittnar and J.~\v{S}ejnoha, \emph{Numerical methods in structural
  mechanics}, ASCE Press and Thomas Telford, Ltd, New York and London, 1996.

\bibitem{Chati:1998:PEP}
M.K. Chati and A.K. Mitra, \emph{Prediction of elastic properties of
  fiber-reinforced unidirectional composites}, Engineering Analysis With
  Boundary Elements \textbf{21} (1998), no.~3, 235--244.

\bibitem{Cox:2006:QVT}
B.~Cox and Q.~Yang, \emph{In quest of virtual tests for structural composites},
  Science \textbf{314} (2006), no.~5802, 1102--1107.

\bibitem{Dobert:2000:NSID}
C.~D\"{o}bert, R.~Mahnken, and E.~Stein, \emph{Numerical simulation of
  interface debonding with a combined damage/friction constitutive model},
  Computational Mechanics \textbf{25} (2000), no.~5, 456--467.

\bibitem{Dong:2000:MFE}
Z.~Dong and A.~J. Levy, \emph{Mean field estimates of the response of fiber
  composites with nonlinear interface}, Mechanics of Materials \textbf{32}
  (2000), no.~12, 739--767.

\bibitem{Dostal:2009:QP}
Z.~Dost\'{a}l, \emph{Optimal quadratic programming algorithms with applications
  to variational inequalities}, Springer Optimization and Its Applications,
  vol.~23, Springer Verlag, New York, NY, USA, 2009.

\bibitem{Dostal:2007:FETI}
Z.~Dost\'{a}l, D.~Hor\'{a}k, and O.~Vlach, \emph{{FETI}-based algorithms for
  modelling of fibrous composite materials with debonding}, Mathematics and
  Computers in Simulation \textbf{76} (2007), no.~1--3, 57--64.

\bibitem{Farhat:1994:OCP}
C.~Farhat, J.~Mandel, and F.~X. Roux, \emph{Optimal convergence properties of
  the {FETI} domain decomposition method}, Computer Methods in Applied
  Mechanics and Engineering \textbf{115} (1994), 365--385.

\bibitem{Farhat:1991:FETI}
C.~Farhat and F.-X. Roux, \emph{A method of finite element tearing and
  interconnecting and its parallel solution algorithm}, International Journal
  for Numerical Methods in Engineering \textbf{32} (1991), no.~6, 1205--1227.

\bibitem{Feyel:2000:FE2}
F.~Feyel and J.-L. Chaboche, \emph{{FE}$^2$ multiscale approach for modelling
  the elastoviscoplastic behaviour of long fibre {SiC/Ti} composite materials},
  Computer Methods in Applied Mechanics and Engineering \textbf{183} (2000),
  no.~3--4, 309--330.

\bibitem{Gosz:1994:RITF}
M.~Gosz, B.~Moran, and J.~D. Achenbach, \emph{On the role of interphases in the
  transverse failure of fiber composites}, International Journal of Damage
  Mechanics \textbf{3} (1994), no.~4, 357--377.

\bibitem{Gruber:2008:FBH}
P.~Gruber, \emph{{FETI}-based homogenization of composites with perfect bonding
  and debonding of constituents}, Bulletin of Applied Mechanics \textbf{4}
  (2008), no.~13, 11--17.

\bibitem{Gruber:2008:HCM}
P.~Gruber, \emph{Homogenization of composite materials with imperfect bonding of
  constituents}, Master's thesis, Czech Technical University in Prague, 2008,
  (in
  Czech)\\\url{http://mech.fsv.cvut.cz/~grubepav/download/gruber_master_thesis%
.pdf}, [January 27, 2009].

\bibitem{Hashin:1990:TPFC}
Z.~Hashin, \emph{Thermoelastic properties of fiber composites with imperfect
  interface}, Mechanics of Materials \textbf{8} (1990), no.~4, 333--348.

\bibitem{Inglis:2007:CMDP}
H.~M. Inglis, P.~H. Geubelle, K.~Matou\v{s}, H.~Tan, and Y.~Huang,
  \emph{Cohesive modeling of dewetting in particulate composites:
  {M}icromechanics vs. multiscale finite element analysis}, Mechanics of
  Materials \textbf{39} (2007), no.~6, 580--595.

\bibitem{Kanit:2003:DSRVE}
T.~Kanit, S.~Forest, I.~Galliet, V.~Mounoury, and D.~Jeulin,
  \emph{Determination of the size of the representative volume element for
  random composites: statistical and numerical approach}, International Journal
  of Solids and Structures \textbf{40} (2003), no.~13--14, 3647--3679.

\bibitem{Kouznetsova:2001:AMMM}
V.~Kouznetsova, W.~A.~M. Brekelmans, and P.~T. Baaijens, \emph{An approach to
  micro-macro modeling of heterogeneous materials}, Computational Mechanics
  \textbf{27} (2001), no.~1, 37--48.

\bibitem{Kruis:2006:DDM}
J.~Kruis, \emph{Domain decomposition methods for distributed computing},
  Saxe-Coburg Publications, Kippen, Stirling, Scotland, 2006.

\bibitem{Kruis:2008:RMI}
J.~Kruis and Z.~Bittnar, \emph{Reinforcement-matrix interaction modeled by
  {FETI} method}, Domain Decomposition Methods in Science and Engineering XVII
  (U.~Langer, M.~Discacciati, D.~E. Keyes, O.~B. Widlund, and W.~Zulehner,
  eds.), Lecture Notes in Computational Science and Engineering, vol.~60,
  Springer Verlag, 2008, pp.~567--573.

\bibitem{Li:2004:DCMM}
S.~Li and S.~Ghosh, \emph{Debonding in composite microstructures with
  morphological variations}, International Journal of Computational Methods
  \textbf{1} (2004), no.~1, 121--149.

\bibitem{Matous:2007:MMSP}
K.~Matou\v{s}, H.~M. Inglis, X.~Gu, D.~Rypl, T.~L. Jackson, and P.~H. Geubelle,
  \emph{Multiscale modeling of solid propellants: From particle packing to
  failure}, Composites Science and Technology \textbf{67} (2007), no.~7--8,
  1694--1708.

\bibitem{Michel:1999:EPC}
J.~C. Michel, H.~Moulinec, and P.~Suquet, \emph{Effective properties of
  composite materials with periodic microstructure: {A} computational
  approach}, Computer Methods in Applied Mechanics and Engineering \textbf{172}
  (1999), no.~1--4, 109--143.

\bibitem{Milton:2002:TC}
G.~W. Milton, \emph{The theory of composites}, Cambridge Monographs on Applied
  and Computational Mathematics, vol.~6, Cambridge University Press, 2002.

\bibitem{Mori:1973:MTM}
T.~Mori and K.~Tanaka, \emph{Average stress in matrix and average elastic
  energy of elastic materials with misfitting inclusions}, Acta Metallurgica
  \textbf{21} (1973), no.~5, 571--573.

\bibitem{Pagano:1990:MIB}
N.~J. Pagano and G.~P. Tandon, \emph{Modeling of imperfect bonding in fiber
  reinforced brittle matrix composites}, Mechanics of Materials \textbf{9}
  (1990), no.~1, 49--64.

\bibitem{Pendhari:2008:APC}
S.~S. Pendhari, T.~Kant, and Y.~M. Desai, \emph{Application of polymer
  composites in civil construction: {A} general review}, Composite Structures
  \textbf{84} (2008), no.~2, 114--124.

\bibitem{Persson:2004:SMG}
P.-O. Persson and G.~Strang, \emph{A simple mesh generator in {MATLAB}}, SIAM
  Review \textbf{46} (2004), no.~2, 329--345.

\bibitem{Castenada:1995:ESD}
P.~Ponte Casta\~{n}eda and J.R. Willis, \emph{The effect of spatial
  distribution on the effective behaviour of composite materials and cracked
  media}, Journal of the Mechanics and Physics of Solids \textbf{43} (1995),
  no.~12, 1919--1951.

\bibitem{Prochazka:2001:HLD}
P.~Proch\'{a}zka, \emph{Homogenization of linear and of debonding composites
  using the {BEM}}, Engineering Analysis with Boundary Elements \textbf{25}
  (2001), no.~9, 753--769.

\bibitem{Prochazka:1995:DDR}
P.~Proch\'{a}zka and M.~\v{S}ejnoha, \emph{Development of debond region of lag
  model}, Computers \& Structures \textbf{55} (1995), no.~2, 249--260.

\bibitem{Rekik:2007:OEL}
A.~Rekik, F.~Auslender, M.~Bornert, and A.~Zaoui, \emph{Objective evaluation of
  linearization procedures in nonlinear homogenization: {A} methodology and
  some implications on the accuracy of micromechanical schemes}, International
  Journal of Solids and Structures \textbf{44} (2007), no.~10, 3468--3496.

\bibitem{rektorys:1999:VMI}
K.~Rektorys, \emph{Variational methods in mathematics, science and
  engineering}, second ed., Springer Verlag, 2007.

\bibitem{Schellekens:1993:NIIE}
J.~C.~J. Schellekens and R.~De~Borst, \emph{On the numerical integration of
  interface elements}, International Journal for Numerical Methods in
  Engineering \textbf{36} (1993), no.~1, 43--66.

\bibitem{Tan:2005:MTM}
H.~Tan, Y.~Huang, C.~Liu, and P.~H. Geubelle, \emph{The {M}ori-{T}anaka method
  for composite materials with nonlinear interface debonding}, International
  Journal of Plasticity \textbf{21} (2005), no.~10, 1890--1918.

\bibitem{Tandon:1996:ETM}
G.~P. Tandon and N.~J. Pagano, \emph{Effective thermoelastic moduli of a
  unidirectional fiber composite containing interfacial arc microcracks},
  Journal of Applied Mechanics \textbf{63} (1996), no.~1, 210--217.

\bibitem{Teng:2007:TSP}
H.~Teng, \emph{Transverse stiffness properties of unidirectional fiber
  composites containing debonded fibers}, Composites Part A: Applied Science
  and Manufacturing \textbf{38} (2007), no.~3, 682--690.

\bibitem{Vlach:2001:MUD}
O.~Vlach, \emph{Modelling of composites using duality-based solvers}, Master's
  thesis, Technical University of Ostrava, 2001, (in Czech)\\
  \url{http://dspace.vsb.cz/handle/10084/41719}, [January 27, 2009].

\bibitem{Sejnoha:1998:MEP}
M.~\v{S}ejnoha and M.~Srinivas, \emph{Modeling of effective properties of
  composites with interfacial microcracks using {PHA} model}, Building Research
  Journal \textbf{46} (1998), no.~2, 99--108.

\bibitem{Sejnoha:2004:NVA}
M.~\v{S}ejnoha, R.~Valenta, and J.~Zeman, \emph{Nonlinear viscoelastic analysis
  of statistically homogeneous random composites}, International Journal for
  Multiscale Computational Engineering \textbf{2} (2004), no.~4, 644--672.

\bibitem{Walpole:1969:OOM}
L.~J. Walpole, \emph{On the overall elastic moduli of composite materials},
  Journal of the Mechanics and Physics of Solids \textbf{17} (1969), no.~4,
  235--251.

\bibitem{Walter:1997:CMD}
M.~E. Walter, G.~Ravichandran, and M.~Ortiz, \emph{Computational modeling of
  damage evolution in unidirectional fiber reinforced ceramic matrix
  composites}, Computational Mechanics \textbf{20} (1997), no.~1, 192--198.

\bibitem{Wriggers:1998:CSID}
P.~Wriggers, G.~Zavarise, and T.~I. Zohdi, \emph{A computational study of
  interfacial debonding damage in fibrous composite materials}, Computational
  Materials Science \textbf{12} (1998), no.~1, 39--56.

\bibitem{Zeman:2001:NEE}
J.~Zeman and M.~\v{S}ejnoha, \emph{Numerical evaluation of effective elastic
  properties of graphite fiber tow impregnated by polymer matrix}, Journal of
  the Mechanics and Physics of Solids \textbf{49} (2001), no.~1, 69--90.

\bibitem{Zeman:2007:FRM}
J.~Zeman and M.~\v{S}ejnoha, \emph{From random microstructures to representative volume elements},
  Modelling and Simulation in Materials Science and Engineering \textbf{15}
  (2007), no.~4, S325--S335.

\end{thebibliography}
\end{document}